\documentclass[10pt, final, journal, letterpaper, oneside, twocolumn]{IEEEtran}
\usepackage{graphicx}
\usepackage{amsmath}
\usepackage{amssymb}
\usepackage{mathrsfs}
\usepackage{stfloats}
\usepackage{dsfont}
\usepackage{setspace}
\begin{document}
\title{Space Shift Keying (SSK--) MIMO with Practical Channel Estimates}
\author{Marco~Di~Renzo,~\IEEEmembership{Member,~IEEE},
        Dario~De~Leonardis,
        Fabio~Graziosi,~\IEEEmembership{Member,~IEEE}, and
        Harald Haas,~\IEEEmembership{Member,~IEEE}
\thanks{Paper approved by A. Zanella, the Editor for Wireless Systems of the IEEE Communications Society. Manuscript received December 18, 2010; revised April 23, 2011; October 9, 2011; and November 16, 2011.}
\thanks{This paper was presented in part at the IEEE Global Communications Conference (GLOBECOM), Houston, TX, USA, December 2011.}
\thanks{M. Di Renzo is with L2S, UMR 8506 CNRS -- SUPELEC -- Univ Paris--Sud, 3 rue Joliot--Curie, 91192 Gif--sur--Yvette CEDEX (Paris), France
(e--mail: marco.direnzo@lss.supelec.fr).}
\thanks{D. De Leonardis and F. Graziosi are with The University of L'Aquila, Department of Electrical and Information Engineering, Center of Excellence for Research DEWS, via G. Gronchi 18, Nucleo Industriale di Pile, 67100 L'Aquila, Italy (e--mail: fabio.graziosi@univaq.it).}
\thanks{H. Haas is with The University of Edinburgh, School of Engineering, Institute for Digital Communications (IDCOM), King's Buildings, Alexander Graham Bell Building, Mayfield Road, Edinburgh, EH9 3JL, Scotland, United Kingdom (UK) (e--mail: h.haas@ed.ac.uk).}
\thanks{Digital Object Identifier XX.XXXX/TCOMM.XXXX.XX.XXXXXX.}}
\markboth{Transactions on Communications} {M. Di Renzo, D. De Leonardis, F. Graziosi, and H. Haas: Space Shift Keying (SSK--) MIMO with
Practical Channel Estimates}
\maketitle
\begin{abstract}
In this paper, we study the performance of space modulation for Multiple--Input--Multiple--Output (MIMO) wireless systems with imperfect
channel knowledge at the receiver. We focus our attention on two transmission technologies, which are the building blocks
of space modulation: i) Space Shift Keying (SSK) modulation; and ii) Time--Orthogonal--Signal--Design (TOSD--) SSK modulation, which is an
improved version of SSK modulation providing transmit--diversity. We develop a single--integral closed--form analytical framework to compute
the Average Bit Error Probability (ABEP) of a mismatched detector for both SSK and TOSD--SSK modulations. The framework exploits the
theory of quadratic--forms in conditional complex Gaussian Random Variables (RVs) along with the Gil--Pelaez inversion theorem. The analytical
model is very general and can be used for arbitrary transmit-- and receive--antennas, fading distributions, fading spatial correlations, and
training pilots. The analytical derivation is substantiated through Monte Carlo simulations, and it is shown, over independent and identically
distributed (i.i.d.) Rayleigh fading channels, that SSK modulation is as robust as single--antenna systems to imperfect channel knowledge, and
that TOSD--SSK modulation is more robust to channel estimation errors than the Alamouti scheme. Furthermore, it is pointed out that only few
training pilots are needed to get reliable enough channel estimates for data detection, and that transmit-- and receive--diversity of SSK
and TOSD--SSK modulations are preserved even with imperfect channel knowledge.
\end{abstract}
\section{Introduction} \label{Intro}
\PARstart{S}{pace} modulation \cite{MDR_Magazine} is a novel digital modulation concept for Multiple--Input--Multiple--Output (MIMO) wireless systems, which is
receiving a growing attention due to the possibility of realizing low--complexity and spectrally--efficient MIMO implementations
\cite{Yang_2008}--\cite{MDR_TCOM2010}. The space modulation principle is known in the literature in various forms, such as Information--Guided
Channel Hopping (IGCH) \cite{Yang_2008}, Spatial Modulation (SM) \cite{Haas_TVT}, and Space Shift Keying (SSK) modulation \cite{Ghrayeb_TWC}.
Although different from one another, all these transmission technologies share the same fundamental working principle, which makes them unique
with respect to conventional modulation schemes: they encode part of the information bits into the spatial positions of the transmit--antennas in the
antenna--array, which plays the role of a constellation diagram (the so--called ``spatial--constellation diagram'') for data modulation \cite{MDR_Magazine},
\cite{MDR_TCOM2010}. In particular, SSK modulation exploits only the spatial--constellation diagram for data modulation, which results in a
very low--complexity modulation concept for MIMO systems \cite{Ghrayeb_TWC}. Recently, improved space modulation schemes that can achieve a
transmit--diversity gain have been proposed in \cite{MDR_TOSD}--\cite{MDR_TOSD_TCOM}. Furthermore, a unified MIMO architecture based on the SSK
modulation principle has been introduced in \cite{Hanzo_VTC2010}.

In SSK modulation, blocks of information bits are mapped into the index of a single transmit--antenna, which is switched on for data
transmission while all the other antennas radiate no power \cite{Ghrayeb_TWC}. SSK modulation exploits the location--specific property of the
wireless channel for data modulation \cite{MDR_TCOM2010}: the messages sent by the transmitter can be decoded at the destination since the
receiver sees a different Channel Impulse Response (CIR) on any transmit--to--receive wireless link. In \cite{Ghrayeb_TWC} and
\cite{MDR_TCOM2010}, it has been shown that the CIRs are the points of the spatial--constellation diagram, and that the Bit Error Probability
(BEP) depends on the distance among these points. Recent results have shown that, if the receiver has Perfect Channel State Information
(P--CSI), space modulation can provide better performance than conventional modulation schemes with similar complexity
\cite{Yang_2008}--\cite{Ghrayeb_TWC}, \cite{Basar_PIMRC2010}, \cite{Hanzo_2010}, and \cite{Ghrayeb_CL}--\cite{MDR_CHNACOM2010}. However, due to
its inherent working principle, the major criticism about the adoption of SSK modulation in realistic propagation environments is its
robustness to the imperfect knowledge of the wireless channel at the receiver. In particular, it is often argued that space modulation is more
sensitive to channel estimation errors than conventional systems. The main contribution of this paper is to shed light on this matter.

Some research works on the performance of space modulation with imperfect channel knowledge are available in the literature. However, they are
insufficient and only based on numerical simulations. In \cite{Ghrayeb_TWC}, the authors have studied the ABEP of SSK modulation with
non--ideal channel knowledge. However, there are four limitations in this paper: i) the ABEP is obtained only through Monte Carlo simulations,
which is not very much insightful; ii) the arguments in \cite{Ghrayeb_TWC} are applicable only to Gaussian fading channels and do not take into
account the cross--product between channel estimation error and Additive White Gaussian Noise (AWGN) at the receiver; iii) it is unclear from
\cite{Ghrayeb_TWC} how the ABEP changes with the pilot symbols used by the channel estimator; and iv) the robustness/weakness of SSK modulation
with respect to conventional modulation schemes is not analyzed. In \cite{MDR_TCOM_PCSI}, we have studied the performance of SSK modulation
when the receiver does not exploit for data detection the knowledge of the phase of the channel gains (semi--blind receiver). It is shown that
semi--blind receivers are much worse than coherent detection schemes, and, thus, that the assessment of the performance of coherent detection
with imperfect channel knowledge is a crucial aspect for SSK modulation. A very interesting study has been recently conducted in
\cite{Ulla_Faiz}, where the authors have compared the performance of SM and V--BLAST (Vertical Bell Laboratories Layered Space--Time)
\cite{V_BLAST} schemes with practical channel estimates. It is shown that the claimed sensitivity of space modulation to channel estimation
errors is simply a misconception and that, on the contrary, SM is more robust than V--BLAST to imperfections on the channel estimates, and that
less training is, in general, needed. However, the study in \cite{Ulla_Faiz} is conducted only through Monte Carlo simulations, which do not
give too much insights for performance analysis and system optimization. Finally, in \cite{Hanzo_2010} the authors have proposed a Differential
Space--Time Shift Keying (DSTSK) scheme, which is based on the Cayley unitary transform theory. The DSTSK scheme requires no channel estimation
at the receiver, but incurs in a 3dB performance loss with respect to coherent detection. Furthermore, it can be applied to only real--valued
signal constellations. Unlike \cite{Hanzo_2010}, which avoids channel estimation, we are interested in studying the training overhead that is
needed for channel estimation and to achieve close--to--optimal performance with coherent detection.

Motivated by these considerations, this paper is aimed at developing a very general analytical framework to assess the performance of space
modulation with coherent detection and practical channel estimates. In particular, we focus our attention on two transmission technologies,
which are the building blocks of space modulation: i) Space Shift Keying (SSK) modulation \cite{Ghrayeb_TWC}; and ii)
Time--Orthogonal--Signal--Design (TOSD--) SSK modulation, which is an improved version of SSK modulation providing transmit--diversity
\cite{MDR_TOSD}, \cite{MDR_TOSD_TCOM}. Our theoretical and numerical results corroborate the findings in \cite{Ulla_Faiz}, and highlight three
important outcomes: i) SSK modulation is as robust as single--antenna systems to imperfect channel knowledge; ii) TOSD--SSK modulation is more
robust to channel estimation errors than the Alamouti scheme \cite{Alamouti}; and iii) only few training pilots are needed to get reliable
enough channel estimates for data detection. More precisely, we provide the following contributions: i) we develop a single--integral
closed--form analytical framework to compute the Average BEP (ABEP) of a mismatched detector \cite{Biglieri} for SSK and TOSD--SSK modulations,
which can be used for arbitrary transmit-- and receive--antennas, fading distributions, fading spatial correlations, and training pilots
for channel estimation. It is shown that the mismatched detector of SSK and TOSD--SSK modulations can be cast in terms of a quadratic--form in
complex Gaussian Random Variables (RVs) when conditioning upon fading channel statistics, and that the ABEP can be computed by exploiting the
Gil--Pelaez inversion theorem \cite{MDR_QF}; ii) over independent and identically distributed (i.i.d.) Rayleigh fading channels, we show that
SSK modulation is superior to Quadrature Amplitude Modulation (QAM), regardless of the number of training pulses, if the spectral efficiency is
greater than 2 bpcu (bits per channel use) and the receiver has at least two antennas; iii) in the same fading channel, we show that TOSD--SSK
modulation is superior, regardless of the number of antennas at the receiver and training pulses, to the Alamouti scheme with QAM if the
spectral efficiency is greater than 2 bpcu. Also, unlike the P--CSI setup, TOSD--SSK modulation can outperform the Alamouti scheme if the
spectral efficiency is 2 bpcu, just one pilot pulse for channel estimation is used, and the detector is equipped with at least two antennas;
iv) still over i.i.d. Rayleigh fading, we show that, compared to the P--CSI scenario, SSK and TOSD--SSK modulations have a
Signal--to--Noise--Ratio (SNR) penalty of approximately 3dB and 2dB when only one pilot pulse can be used for channel estimation, respectively.
Also, single--antenna and Alamouti schemes have a SNR penalty of approximately 3dB for QAM; and v) we verify that transmit--
and receive--diversity of SSK and TOSD--SSK modulations are preserved even for a mismatched detector.

The remainder of this paper is organized as follows. In Section \ref{System_Model}, the system model is introduced. In Section \ref{SSK} and
Section \ref{TOSD-SSK}, SSK and TOSD--SSK modulations are described and the analytical frameworks to compute the ABEP with imperfect channel
knowledge are developed, respectively. In Section \ref{BandwidthPulses}, the spectral efficiency of TOSD--SSK modulation with time--orthogonal
shaping filters is studied. In Section \ref{Results}, numerical results are shown to substantiate the analytical derivation, and to compare SSK and
TOSD--SSK modulations with state--of--the--art single--antenna and Alamouti schemes. Finally, Section \ref{Conclusion} concludes
this paper.
\section{System Model} \label{System_Model}
We consider a generic $N_t \times N_r$ MIMO system, with $N_t$ and $N_r$ being the number of antennas at the transmitter and at the receiver,
respectively. SSK and TOSD--SSK modulations work as follows \cite{Ghrayeb_TWC}, \cite{MDR_TOSD}, \cite{MDR_TOSD_TCOM}: i) the transmitter
encodes blocks of $\log _2 \left(  N_t \right)$ data bits into the index of a single transmit--antenna, which is switched on for data
transmission while all the other antennas are kept silent; and ii) the receiver solves an $N_t$--hypothesis detection problem to estimate the
transmit--antenna that is not idle, which results in the estimation of the unique sequence of bits emitted by the encoder. With respect to SSK
modulation \cite{Ghrayeb_TWC}, in TOSD--SSK modulation \cite{MDR_TOSD_TCOM} the $t$--th transmit--antenna, when active, radiates a distinct
pulse waveform ${w_t \left(  \cdot \right)}$ for $t=1, 2,\ldots, N_t$, and the waveforms across the antennas are time--orthogonal,
\emph{i.e.}\footnote{$\left(  \cdot \right)^*$ denotes complex--conjugate.}, $\int\nolimits_{ - \infty }^{ + \infty } {w_{t_1} \left( \xi
\right)w_{t_2}^* \left( \xi \right)d\xi}  = 0$ if $t_1 \ne t_2$ and $\int\nolimits_{ - \infty }^{ + \infty } {w_{t_1} \left( \xi
\right)w_{t_2}^* \left( \xi \right)d\xi} = 1$ if $t_1 = t_2$. In other words, SSK modulation is a special case of TOSD--SSK modulation with
$w_{t} \left( \xi \right) = w_0 \left( \xi \right)$ for $t=1, 2,\ldots, N_t$.

In \cite{MDR_TOSD_TCOM}, we have analytically proved that the diversity order of SSK modulation is $N_r$, while the diversity order of
TOSD--SSK modulation is $2N_r$, which results in a transmit--diversity equal to 2 and a receive--diversity equal to $N_r$. Thus, TOSD--SSK
modulation provides a full--diversity--achieving (\emph{i.e.}, the diversity gain is $N_t N_r$) system if $N_t=2$. This scheme has been
recently generalized in \cite{MDR_TOSD_ICC2011} to achieve arbitrary transmit--diversity. It is worth emphasizing that in TOSD--SSK modulation
a single--antenna is active for data transmission and that the information bits are still encoded into the index of the transmit--antenna, and
are \emph{not} encoded into the impulse (time) response, ${w_t \left(  \cdot \right)}$, of the shaping filter. In other words, TOSD--SSK
modulation is different from conventional Single--Input--Single--Output (SISO) schemes with Orthogonal Pulse Shape Modulation (O--PSM)
\cite{TCOM_PSM}, which are unable to achieve transmit--diversity as only a single wireless link is exploited for communication
\cite{MDR_TOSD_TCOM}. Also, TOSD--SSK modulation is different from conventional transmit--diversity schemes \cite{TD_CommMag}, and requires no
extra time--slots for transmit--diversity. Further details are available in \cite{MDR_TOSD_TCOM} and are here omitted to avoid repetitions.

Throughout this paper, the block of information bits encoded into the index of the $t$--th transmit--antenna is called ``message'', and it is
denoted by $m_t$ for $t=1, 2,\ldots, N_t$. The $N_t$ messages are assumed to be equiprobable. Moreover, the related transmitted signal is
denoted by $s_t \left( \cdot \right)$. It is implicitly assumed with this notation that, if $m_t$ is transmitted, the analog signal $s_t \left(
\cdot \right)$ is emitted by the $t$--th transmit--antenna while the other antennas radiate no power.
\subsection{Notation} \label{Notation}
Main notation is as follows. i) We adopt a complex--envelope signal representation. ii) $j = \sqrt { - 1}$ is the
imaginary unit. iii) $\left( {x \otimes y} \right)\left( u \right) = \int_{ - \infty }^{ + \infty } {x\left( \xi  \right)y\left( {u - \xi }
\right)d\xi }$ is the convolution of signals $x\left(  \cdot \right)$ and $y\left(  \cdot  \right)$. iv) $\left| {\cdot} \right|^2$ is the
square absolute value. v) $\textrm{E}\left\{ \cdot  \right\}$ is the expectation operator computed over channel fading statistics. vi)
${\mathop{\rm Re}\nolimits} \left\{ \cdot \right\}$ and ${\mathop{\rm Im}\nolimits} \left\{ \cdot \right\}$ are real and imaginary part
operators, respectively. vii) $\Pr \left\{ \cdot \right\}$ denotes probability. viii) $Q\left( u \right) = \left( {{1 \mathord{\left/
{\vphantom {1 {\sqrt {2\pi } }}} \right. \kern-\nulldelimiterspace} {\sqrt {2\pi } }}} \right)\int_u^{ + \infty } {\exp \left( { - {{\xi^2 }
\mathord{\left/ {\vphantom {{\xi^2 } 2}} \right. \kern-\nulldelimiterspace} 2}} \right)d\xi}$ is the Q--function. ix) $\delta \left(  \cdot
\right)$ and $\delta _{ \cdot , \cdot }$ are Dirac and Kronecker delta functions, respectively. x) $M_X \left( s \right) = \textrm{E}\left\{
{\exp \left( { sX} \right)} \right\}$ and $\Psi _X \left( \nu  \right) = \textrm{E}\left\{ {\exp \left( { j\nu X} \right)} \right\}$ are
Moment Generating Function (MGF) and Characteristic Function (CF) of RV $X$, respectively. xi) $\propto$ denotes ``is proportional to''.
\subsection{Channel Model} \label{Channel_Model}
We consider a general frequency--flat slowly--varying channel model with generically correlated and non--identically distributed fading gains.
In particular ($t=1, 2, \ldots, N_t$, $r=1, 2, \ldots, N_r$):
\begin{itemize}
\item $h_{t,r} \left( \xi \right) = \alpha _{t,r} \delta \left( {\xi - \tau _{t,r} } \right)$ is the channel impulse response of the
transmit--to--receive wireless link from the $t$--th transmit--antenna to the $r$--th receive--antenna. $\alpha _{t,r}  = \beta _{t,r} \exp
\left( {j\varphi _{t,r} } \right)$ is the complex channel gain with $\beta _{t,r}$ and $\varphi _{t,r}$ denoting the channel envelope and
phase, respectively, and $\tau _{t,r}$ is the propagation time--delay.
\item The time--delays $\tau _{t,r}$ are assumed to be known at the receiver, \emph{i.e.},
perfect time--synchronization is considered. Furthermore, we consider $\tau _{1,1} \cong \tau _{1,2} \cong \ldots \cong \tau _{N_t, N_r }$, which
is a realistic assumption when the distance between the transmitter and the receiver is much larger than the spacing between the transmit-- and
receive--antennas \cite{MDR_TCOM2010}. Due to the these assumptions, the propagation delays can be neglected in the remainder of this paper.
\end{itemize}
\subsection{Channel Estimation} \label{Channel_Estimation}
Let $E_p$ and $N_p$ be the energy transmitted for each pilot pulse and the number of pilot pulses used for channel estimation, respectively.
Similar to \cite{Proakis_1968} and \cite{Gifford_2005}, we assume that channel estimation is performed by using a Maximum--Likelihood (ML)
detector, and by observing $N_p$ pilot pulses that are transmitted before the modulated data. During the transmission of one block of
pilot--plus--data symbols, the wireless channel is assumed to be constant, \emph{i.e.} a quasi--static channel model is considered. With these
assumptions, the estimates of channel gains $\alpha _{t,r}$ ($t=1, 2, \ldots, N_t$, $r=1, 2, \ldots, N_r$) can be written as follows:
\setcounter{equation}{0}
\begin{equation}
\label{Eq_1} \hat \alpha _{t,r}  = \hat \beta _{t,r} \exp \left( {j\hat \varphi _{t,r} } \right) = \alpha _{t,r}  + \varepsilon _{t,r}
\end{equation}
\noindent where $\hat \alpha _{t,r}$, $\hat \beta _{t,r}$, and ${\hat \varphi _{t,r} }$ are the estimates of $\alpha _{t,r}$, $\beta _{t,r}$,
and ${\varphi _{t,r} }$, respectively, at the output of the channel estimation unit, and $\varepsilon _{t,r}$ is the additive channel
estimation error, which can be shown to be complex Gaussian distributed with zero--mean and variance $\sigma _\varepsilon ^2  = {{N_0 }
\mathord{\left/ {\vphantom {{N_0 } {\left( {E_p N_p } \right)}}} \right. \kern-\nulldelimiterspace} {\left( {E_p N_p } \right)}}$ per dimension
\cite{Proakis_1968}, \cite{Gifford_2005}, where $N_0$ denotes the power spectral density per dimension of the AWGN at the receiver. The channel
estimation errors, $\varepsilon _{t,r}$, are statistically independent and identically distributed, as well as statistically independent of the
channel gains and the AWGN at the receiver.
\begin{figure*}[!t]
\setcounter{equation}{3}
\begin{equation}
\label{Eq_4} \hat D_{m_q } \left( {m_t } \right) = -\sum\limits_{r = 1}^{N_r } {\left[ {\int_{T_m } {\left| {z_r \left( \xi \right) - \hat
s_{t,r} \left( \xi \right)} \right|^2 d\xi} } \right]} \propto \sum\limits_{r = 1}^{N_r } {\left[ {{\mathop{\rm Re}\limits} \left\{
{\int\nolimits_{T_m } {z_r \left( \xi \right)\hat s_{t,r}^ * \left( \xi \right)d\xi} } \right\} - \frac{1}{2}\int\nolimits_{T_m } {\hat s_{t,r}
\left( \xi \right)\hat s_{t,r}^ * \left( \xi \right)d\xi} } \right]}
\end{equation}
\normalsize \hrulefill \vspace*{-6pt}
\end{figure*}
\begin{figure*}[!t]
\setcounter{equation}{4}
\begin{equation}
\label{Eq_5} \hat D_{m_q } \left( {m_t } \right)  =  - \sum\limits_{r = 1}^{N_r } {\left\{ {\int_{T_m } {\left| {\left[ {\sqrt {\frac{{E_m
}}{{N_0 }}} \alpha _{q,r} w_0 \left( \xi \right) + \frac{{\eta _r \left( \xi \right)}}{{\sqrt {N_0 } }}} \right] - \left[ {\sqrt {\frac{{E_m
}}{{N_0 }}} \alpha _{t,r} w_0 \left( \xi \right) + \sqrt {\frac{{E_m }}{{N_0 }}} \varepsilon _{t,r} w_0 \left( \xi \right)} \right]} \right|^2
d\xi} } \right\}}
\end{equation}
\normalsize \hrulefill \vspace*{-10pt}
\end{figure*}
\begin{figure*}[!t]
\setcounter{equation}{5}
\begin{equation}
\label{Eq_6}
\begin{split}
\hat m  &= \mathop {\arg \max }\limits_{m_t \;{\rm{for}}\;t = 1,2, \ldots ,N_t } \left\{ {\hat D_{m_q } \left( {m_t } \right)} \right\}
  \propto \mathop {\arg \min }\limits_{m_t \;{\rm{for}}\;t = 1,2, \ldots ,N_t } \left\{ {\hat D_{m_q }^{\left( e \right)} \left( {m_t } \right)} \right\} \\ & =  \mathop {\arg \min }\limits_{m_t \;{\rm{for}}\;t = 1,2, \ldots ,N_t } \left\{ {\sum\limits_{r = 1}^{N_r } { {\left| {\frac{{\tilde \eta _{0,r} }}{{\sqrt {N_0 } }} - \left[ {\sqrt {\frac{{E_m }}{{N_0 }}} \left( {\alpha _{t,r}  - \alpha _{q,r} } \right) + \sqrt {\frac{{E_m }}{{N_0 }}} \varepsilon _{t,r} } \right]} \right|^2 } } } \right\} \\
 \end{split}
\end{equation}
\normalsize \hrulefill \vspace*{-10pt}
\end{figure*}
\subsection{Mismatched ML--Optimum Detector} \label{ML_Detector}
For data detection, we consider the so--called mismatched ML--optimum receiver according to the definition given in \cite{Biglieri}. In
particular, a detector with mismatched metric estimates the complex channel gains as in (\ref{Eq_1}), and uses them in the same metric that
would be applied if the channels were perfectly known. To avoid repetitions in the analysis of SSK and TOSD--SSK modulations, the mismatched detector is here described by assuming arbitrary shaping filters.

The mismatched ML--optimum detector can be obtained as follows. Let $m_q$ with $q=1, 2,\ldots, N_t$ be the transmitted message. The signal received after
propagation through the wireless fading channel and impinging upon the $r$--th receive--antenna can be written as follows:
\setcounter{equation}{1}
\begin{equation}
\label{Eq_2} z_r\left( \xi \right) = \tilde s_{q,r} \left( \xi \right) + \eta_r\left( \xi \right)\quad \quad {\rm{if}}\;m_q
\;{\rm{is}}\;{\rm{sent}}
\end{equation}
\noindent where: i) $\tilde s_{q,r} \left( \xi \right) = \left( {s_q  \otimes h_{q,r} } \right)\left( \xi \right) = \alpha _{q,r} s_q \left(
{\xi} \right) = \beta _{q,r} \exp \left( {j\varphi _{q,r} } \right)s_q \left( {\xi} \right)$ for $q=1, 2,\ldots, N_t$ and $r=1, 2,\ldots, N_r$;
ii) $s_q \left( \xi \right) = \sqrt {E_m } w_q \left( \xi \right)$ for $q=1, 2,\ldots, N_t$, where $E_m$ is the average energy transmitted by
each antenna that emits a non--zero signal; and iii) $\eta_r\left( \cdot \right)$ is the complex AWGN at the input of the $r$--th
receive--antenna for $r=1,2,\ldots,N_r$, which has power spectral density $N_0$ per dimension. Across the receive--antennas, the noises
$\eta_r\left( \cdot \right)$ are statistically independent.

In particular, (\ref{Eq_2}) is a general $N_t$--hypothesis detection problem  \cite[Sec. 7.1]{Simon}, \cite[Sec. 4.2, pp. 257]{VanTrees} in
AWGN, when conditioning upon fading channel statistics. Accordingly, the mismatched ML--optimum detector with imperfect CSI at the receiver is
as follows:
\setcounter{equation}{2}
\begin{equation}
\label{Eq_3} \hat m = \mathop {\arg \max }\limits_{m_t \; {\rm{ for }} \; t = 1,2, \ldots ,N_t } \left\{ {\hat D_{m_q } \left( {m_t } \right)}
\right\}
\end{equation}
\noindent where $\hat m$ is the estimated message and $\hat D_{m_q } \left( {m_t } \right)$ is the mismatched decision
metric \cite{MDR_TCOM2010}, \cite{MDR_TOSD_TCOM}, which is shown in (\ref{Eq_4}) on top of this page, where $\hat s_{t,r} \left( \xi \right) = \hat \alpha _{t,r} s_t \left( \xi \right) = \left( {\alpha _{t,r}  + \varepsilon _{t,r} } \right)s_t \left( \xi \right)$ and  $T_m$ is the symbol period.
\section{SSK Modulation} \label{SSK}
\subsection{Decision Metrics} \label{DecisionMetrics__SSK}
In SSK modulation, the decision metric in (\ref{Eq_4}) can be re--written from (\ref{Eq_1}) and (\ref{Eq_2}) as shown in (\ref{Eq_5}) on top of this page, where we have taken into account that for SSK modulation the shaping filters are all equal to $w_0 \left( \cdot \right)$, and we have introduced the scaling factor ${1 \mathord{\left/ {\vphantom {1 {N_0}}} \right. \kern-\nulldelimiterspace} {N_0 }}$, which does not affect (\ref{Eq_3}).

From (\ref{Eq_5}), and after some algebra, the maximization problem in (\ref{Eq_3}) reduces to (\ref{Eq_6}) shown on top of this page, where $\tilde \eta_{0,r}  = \int\nolimits_{T_m } {\eta_r \left( \xi \right)w_0^ *  \left( \xi \right)d\xi}$, and ${\hat D_{m_q}^{\left( e \right)} \left( {m_t } \right)}$ is statistically equivalent to $\hat D_{m_q } \left( {m_t } \right)$. In particular, (\ref{Eq_6}) can be thought as a mismatched detector in which: i) first, pulse--matched filtering is performed; and ii) then, ML--optimum decoding is applied to the resulting signal.
\subsection{ABEP} \label{ABEP__SSK}
The ABEP of the detector in (\ref{Eq_6}) can be computed in closed--form as follows:
\setcounter{equation}{6}
\begin{equation}
\label{Eq_7}
\begin{split}
 {\rm{ABEP}} & \mathop  = \limits^{\left( a \right)} {\rm{E}}\left\{ {\sum\limits_{q = 1}^{N_t } {\sum\limits_{t = 1}^{N_t } {\frac{{N_H \left( {t,q} \right)}}{{N_t \log _2 \left( {N_t } \right)}}\Pr \left\{ {\left. {\hat m = m_t } \right|m_q } \right\}} } } \right\} \\
 & \mathop  \le \limits^{\left( b \right)} \sum\limits_{q = 1}^{N_t } {\sum\limits_{t = 1}^{N_t } {\frac{{N_H \left( {t,q} \right)}}{{N_t \log _2 \left( {N_t } \right)}}\underbrace {{\rm{E}}\left\{ {\Pr \left\{ {m_q  \to m_t } \right\}} \right\}}_{{\rm{APEP}}\left( {m_q  \to m_t } \right)}} }  \\
 \end{split}
\end{equation}
\noindent where $\mathop  = \limits^{\left( a \right)}$ comes from \cite[Eq. (4) and Eq. (5)]{ABEP_General}, and $\mathop  = \limits^{\left( b
\right)}$ is the asymptotically--tight union--bound recently introduced in \cite[Eq. (35)]{MDR_TCOM2010}. Furthermore, ${N_H \left( {t ,q }
\right)}$ is the Hamming distance between the bit--to--antenna--index mappings of ${m_{t } }$ and ${m_{q } }$; and ${\rm{APEP}}\left( {m_{q }
\to m_{t } } \right) = {\rm{E}}\left\{ {{\rm{PEP}}\left( {m_{q }  \to m_{t } } \right)} \right\} = {\rm{E}}\left\{ {\Pr \left\{ {m_{q } \to
m_{t } } \right\}} \right\}$ is the Average Pairwise Error Probability (APEP), \emph{i.e.}, the probability of estimating ${m_{t } }$ when,
instead, ${m_{q } }$ is transmitted, under the assumption that ${m_{t } }$ and ${m_{q } }$ are the only two messages possibly being
transmitted.

Let us note that (\ref{Eq_7}) simplifies significantly when ${\rm{APEP}}\left( {m_{q }  \to m_{t } } \right) = {\rm{APEP}}_{\rm{0}}$
for $t = 1,2, \ldots ,N_t$ and $q  = 1,2, \ldots ,N_t$ (\emph{e.g.}, for i.i.d. fading). In this case, the
ABEP in (\ref{Eq_7}) becomes:
\setcounter{equation}{7}
\begin{equation}
\label{Eq_8} {\rm{ABEP}} \le \frac{{{\rm{APEP}}_0 }}{{N_t \log _2 \left( {N_t } \right)}}\sum\limits_{q  = 1}^{N_t } {\sum\limits_{t  = 1}^{N_t
} {N_H \left( {t ,q } \right)} } \mathop  = \limits^{\left( a \right)} \frac{{N_t }}{2}{\rm{APEP}}_0
\end{equation}
\noindent where $\mathop  = \limits^{\left( a \right)}$ comes from the identity $\sum\nolimits_{q  = 1}^{N_t } {\sum\nolimits_{t  = 1}^{N_t }
{N_H \left( {t ,q } \right)} }  = \left( {{{N_t^2 } \mathord{\left/ {\vphantom {{N_t^2 } 2}} \right. \kern-\nulldelimiterspace} 2}} \right)\log
_2 \left( {N_t } \right)$, which can be derived via direct inspection for all possible bit--to--antenna--index mappings.
\begin{figure*}[!t]
\setcounter{equation}{14}
\begin{equation}
\label{Eq_13} \Psi _D \left( {\left. \nu  \right|{\boldsymbol{\alpha}}_{t,q}} \right) = \frac{{\left( {v_a v_b } \right)^{N_r } }}{{\left( {\nu
+ jv_a } \right)^{N_r } \left( {\nu  - jv_b } \right)^{N_r } }}\exp \left\{ {\frac{{v_a v_b \left( { - \nu ^2 g_a \gamma_{t,q}  + j\nu g_b
\gamma_{t,q} } \right)}}{{\left( {\nu  + jv_a } \right)\left( {\nu  - jv_b } \right)}}} \right\} = \Upsilon \left( \nu  \right)\exp \left\{
{\Delta \left( \nu \right)\gamma_{t,q} } \right\}
\end{equation}
\normalsize \hrulefill \vspace*{-5pt}
\end{figure*}
\begin{figure*}[!t]
\setcounter{equation}{18}
\begin{equation}
\label{Eq_17} {\rm{ABEP}} \le \frac{1}{{N_t \log _2 \left( {N_t } \right)}}\sum\limits_{q = 1}^{N_t } {\sum\limits_{t = 1}^{N_t } {\left[ {N_H
\left( {t,q} \right)\left( {\frac{1}{2} - \frac{1}{\pi }\int\nolimits_0^{ + \infty } { {\frac{{{\mathop{\rm Im}\nolimits} \left\{
{\Upsilon \left( \nu  \right)M_{\gamma _{t,q} } \left( {\Delta \left( \nu  \right)} \right)} \right\}}}{\nu }} d\nu } } \right)}
\right]} }
\end{equation}
\normalsize \hrulefill \vspace*{-10pt}
\end{figure*}
\subsection{Computation of PEPs} \label{PEP__SSK}
Let us start by computing the PEPs, \emph{i.e.}, the pairwise probabilities in (\ref{Eq_7}) when conditioning upon fading channel statistics.
From (\ref{Eq_6}), ${\rm{PEP}}\left( {m_q  \to m_t } \right)$ is as follows:
\setcounter{equation}{8}
\begin{equation}
\label{Eq_9}
 {\rm{PEP}}\left( {m_q  \to m_t } \right) = \Pr \left\{ {\hat D_{m_q }^{\left( e \right)} \left( {m_t } \right) < \hat D_{m_q }^{\left( e \right)} \left( {m_q } \right)} \right\}
\end{equation}
\noindent where:
\setcounter{equation}{9}
\begin{equation}
\label{Eq_9a}
\left\{ \begin{array}{l}
 \hat D_{m_q }^{\left( e \right)} \left( {m_t } \right) = \sum\limits_{r = 1}^{N_r } {\left| {\frac{{\tilde \eta _{0,r} }}{{\sqrt {N_0 } }} - \left[ {\sqrt {\frac{{E_m }}{{N_0 }}} \left( {\alpha _{t,r}  - \alpha _{q,r} } \right) + \sqrt {\frac{{E_m }}{{N_0 }}} \varepsilon _{t,r} } \right]} \right|^2 }  \\
 \hat D_{m_q }^{\left( e \right)} \left( {m_q } \right) = \sum\limits_{r = 1}^{N_r } {\left| {\frac{{\tilde \eta _{0,r} }}{{\sqrt {N_0 } }} - \sqrt {\frac{{E_m }}{{N_0 }}} \varepsilon _{q,r} } \right|^2 }  \\
 \end{array} \right.
\end{equation}

By introducing the notation ($r=1,2,\ldots,N_r$):
\setcounter{equation}{10}
\begin{equation}
\label{Eq_10} \left\{ \begin{array}{l}
 X_r  = \frac{{\tilde \eta _{0,r} }}{{\sqrt {N_0 } }} - \left[ {\sqrt {\frac{{E_m }}{{N_0 }}} \left( {\alpha _{t,r}  - \alpha _{q,r} } \right) + \sqrt {\frac{{E_m }}{{N_0 }}} \varepsilon _{t,r} } \right] \\
 Y_r  = \frac{{\tilde \eta _{0,r} }}{{\sqrt {N_0 } }} - \sqrt {\frac{{E_m }}{{N_0 }}} \varepsilon _{q,r}  \\
 \end{array} \right.
\end{equation}
\noindent the PEP in (\ref{Eq_9}) can be re--written in the general form (with $A = 1$, $B = -1$, and $C = 0$):
\setcounter{equation}{11}
\begin{equation}
\label{Eq_11} {\rm{PEP}}\left( {m_q  \to m_t } \right) = \Pr \left\{ {D < 0} \right\}
\end{equation}
\noindent where:
\setcounter{equation}{12}
\begin{equation}
\label{Eq_11a} D = \sum\limits_{r = 1}^{N_r } {\left( {A\left| {X_r } \right|^2  + B\left|
{Y_r } \right|^2  + CX_r Y_r^ *   + C^ *  X_r^ *  Y_r } \right)}
\end{equation}

From \cite[Sec. III]{MDR_QF}, we notice that, when conditioning upon fading channel statistics, the RV $D$ in (\ref{Eq_11a}) is a
quadratic--form in complex Gaussian RVs. In fact, AWGN at the receiver input and channel estimation error are Gaussian distributed
RVs. Furthermore, they are mutually independent among themselves and across the $N_r$ receive--antennas. Literature on quadratic--forms in
complex Gaussian RVs is very rich, and during the last decades many different techniques have been developed for their analysis (see,
\emph{e.g.}, \cite{MDR_QF} and \cite{ProakisQF_1968} for a survey). Furthermore, effective methods for the computation of the PEP over generalized fading channels have been proposed, \emph{e.g.}, \cite{SimonQF}--\cite{Driscoll}, and simple analytical frameworks for some special fading scenarios are available in \cite[Ch. 9]{SimonBookQF}. In this paper, we propose to use the Gil--Pelaez inversion theorem \cite{GilPelaez}.

Accordingly, by using \cite{GilPelaez} the PEP can be computed as follows:
\setcounter{equation}{13}
\begin{equation} \label{Eq_12} \begin{split}
{\rm{PEP}}\left( {m_q  \to m_t } \right) &= \frac{1}{2} - \frac{1}{\pi }\int\nolimits_0^{ + \infty } { {\frac{{{\mathop{\rm
Im}\nolimits} \left\{ {\Psi _D \left( {\left. \nu  \right|{\boldsymbol{\alpha}}_{t,q} } \right)} \right\}}}{\nu }d\nu } } \\ &= \frac{1}{2}
- \frac{1}{\pi }\int\nolimits_0^{{\pi  \mathord{\left/ {\vphantom {\pi  2}} \right. \kern-\nulldelimiterspace} 2}} {{\frac{{{\mathop{\rm Im}\nolimits} \left\{ {\Psi _D \left( {\left. {\tan \left( \xi  \right)} \right|{\boldsymbol{\alpha}}_{t,q} } \right)} \right\}}}{{\sin \left( \xi  \right)\cos \left( \xi  \right)}}d\xi } } \end{split}
\end{equation}
\noindent where $\Psi _D \left( {\left.  \cdot  \right|{\boldsymbol{\alpha}}_{t,q}} \right)$ is the CF of RV $D$ when conditioning upon the
channel gains, and ${\boldsymbol{\alpha}}_{t,q} = \left\{ {\alpha _{t,r} ,\alpha _{q,r} } \right\}_{r = 1}^{N_r }$ is a short--hand to denote
all the channel gains in (\ref{Eq_11a}).

The conditional CF, $\Psi _D \left( {\left.  \cdot  \right|{\boldsymbol{\alpha}}_{t,q}} \right)$, of RV $D$ is given by \cite[Eq. (2) and Eq.
(3)]{MDR_QF} shown in (\ref{Eq_13}) on top of this page, where: i) $\bar \gamma {\rm{ = }}{{E_m } \mathord{\left/ {\vphantom {{E_m } { {N_0 }}}} \right. \kern-\nulldelimiterspace} {{N_0 }}}$; ii) $r_{pm} = {{E_p } \mathord{\left/ {\vphantom {{E_p } {E_m }}} \right. \kern-\nulldelimiterspace} {E_m }}$; iii) $g_a  = 2\bar \gamma \left[ {1 + \left( {N_p r_{pm} } \right)^{ - 1} } \right]$; iv) $g_b  = \bar \gamma$; and:
\setcounter{equation}{15}
\begin{equation} \label{Eq_14}
\left\{ \begin{array}{l}
 v_a  = v_b  = \left( {{1 \mathord{\left/ {\vphantom {1 2}} \right. \kern-\nulldelimiterspace} 2}} \right)\sqrt {\left[ {\left( {N_p r_{pm} } \right)^{ - 2}  + 2\left( {N_p r_{pm} } \right)^{ - 1} } \right]^{ - 1} } \\
 \gamma_{t,q} = \gamma \left( {\boldsymbol{\alpha}}_{t,q} \right) = \sum\limits_{r = 1}^{N_r } {\left| {\alpha _{q,r}  - \alpha _{t,r} } \right|^2 } \\
 \Delta \left( \nu  \right) = v_a v_b \left( { - \nu ^2 g_a  + j\nu g_b } \right)\left( {\nu  + jv_a } \right)^{ - 1} \left( {\nu  - jv_b } \right)^{ - 1} \\
 \Upsilon \left( \nu  \right) = \left( {v_a v_b } \right)^{N_r } \left( {\nu  + jv_a } \right)^{ - N_r } \left( {\nu  - jv_b } \right)^{ - N_r }
 \end{array} \right.
\end{equation}
\subsection{Computation of APEPs} \label{APEP__SSK}
The APEP can be computed from (\ref{Eq_12}) by removing the conditioning over the fading channel:
\setcounter{equation}{16}
\begin{equation}
\label{Eq_15} \begin{split} {\rm{APEP}}\left( {m_q  \to m_t } \right) & = {\rm{E}}\left\{ {{\rm{PEP}}\left( {m_q  \to m_t } \right)} \right\} \\ &= \frac{1}{2} -
\frac{1}{\pi }\int\nolimits_0^{ + \infty } { {\frac{{{\mathop{\rm Im}\nolimits} \left\{ {\Psi _D \left( \nu  \right)} \right\}}}{\nu }} d\nu } \end{split} \end{equation}
\noindent where $\Psi _D \left( \nu  \right) = {\rm{E}}\left\{ {\Psi _D \left( {\left. \nu  \right|{\boldsymbol{\alpha}}_{t,q} } \right)}
\right\}$ is the CF of RV $D$ averaged over all fading channel statistics. It can be computed from (\ref{Eq_13}), as follows:
\setcounter{equation}{17}
\begin{equation}
\label{Eq_16} \Psi _D \left( \nu  \right) = {\rm{E}}\left\{ {\Upsilon \left( \nu  \right)\exp \left\{ {\Delta \left( \nu  \right)\gamma_{t,q}}
\right\}} \right\} \mathop  = \limits^{\left( a \right)} \Upsilon \left( \nu  \right)M_{\gamma_{t,q}}  \left( {\Delta \left( \nu  \right)}
\right)
\end{equation}
\noindent where $M_{\gamma_{t,q}}  \left(  \cdot  \right)$ is the MGF of RV $\gamma_{t,q}$, and $\mathop  =
\limits^{\left( a \right)}$ comes from the definition of MGF.

In conclusion, the ABEP of SSK modulation over arbitrary fading channels and with practical channel estimates can be computed in closed--form
from (\ref{Eq_7}), (\ref{Eq_15}), and (\ref{Eq_16}), as shown in (\ref{Eq_17}) on top of this page.

The formula in (\ref{Eq_17}) provides a very simple analytical tool for performance assessment of SSK modulation with channel estimation
errors, and allows us to estimate the number of pilot pulses, $N_p$, and the fraction of energy, $r_{pm}$, to be allocated to each pilot pulse
to get the desired performance. In particular, (\ref{Eq_17}) needs only the MGF of RV $\gamma_{t,q}$ to be computed. This latter MGF is the
building block for computing the ABEP with P--CSI, and it has been recently computed in closed--form for a number of MIMO setups and fading
conditions. In particular: i) it is known in closed--form for arbitrary correlated Nakagami--\emph{m} fading channels and $N_r=1$
\cite{MDR_TCOM2010}; ii) it can be derived from \cite{MDR_TCOM2010} for independent Nakagami--\emph{m} fading channels and arbitrary
$N_r$ \cite{MDR_Letter2011}; iii) it can be derived from \cite{MDR_TCOM2010} for Nakagami--\emph{m} fading channels and arbitrary $N_r$ when the channel
gains are correlated at the transmitter--side but are independent at the receiver--side \cite{MDR_Letter2011}; and iv) it is known in closed--form for arbitrary
correlated Rician fading channels and arbitrary $N_r$ \cite{MDR_TOSD_TCOM}. For example, for i.i.d. Rayleigh fading channels, the ABEP in
(\ref{Eq_17}) reduces, from (\ref{Eq_8}), to:
\setcounter{equation}{19}
\begin{equation}
\label{Eq_18} {\rm{ABEP}} \le \frac{{N_t }}{4} - \frac{{N_t }}{{2\pi }}\int\nolimits_0^{ + \infty } { {{\mathop{\rm Im}\nolimits} \left\{
{\frac{{\Upsilon \left( \nu  \right)}}{\nu }\frac{1}{{\left( {1 - 2\Omega _0 \Delta \left( \nu  \right)} \right)^{N_r } }}} \right\}} d\nu }
\end{equation}
\noindent where $\Omega _0  = {\rm{E}}\left\{ {\left| {\alpha _{t,r} } \right|^2 } \right\}$ is the mean square value of the i.i.d. channel gains.

Finally, we conclude this section with three general comments about (\ref{Eq_17}): i) the integrand function is, in general, well--behaved when
$\nu  \to 0$ for typical MGFs used in wireless communication problems. Thus, the numerical computation of the integral does not provide any
critical issues. The interested reader might check this out in (\ref{Eq_18}), where it can be shown that the integrand function tends to a
finite value when $\nu  \to 0$; ii) since the ABEP depends on the MGF of RV $\gamma_{t,q}$, from \cite{MDR_TOSD_TCOM} and \cite{Giannakis} we
conclude that the diversity order of the system is given by $N_r$, which is the same as the P--CSI scenario. We will verify this statement in
Section \ref{Results} with some numerical examples, which will highlight that there is no loss in the diversity order with practical channel
estimation; and iii) by direct inspection, it can be shown that (\ref{Eq_17}) reduces to the P--CSI lower--bound if $N_p r_{pm} \to  + \infty$.
\begin{figure*}[!t]
\setcounter{equation}{21}
\begin{equation}
\label{Eq_20} \left\{ \begin{array}{l}
 \hat D_{m_q } \left( {m_t } \right) = \sum\limits_{r = 1}^{N_r } {{\mathop{\rm Re}\nolimits} \left\{ {\left( {\alpha _{t,r}  + \varepsilon _{t,r} } \right)^ *  \sqrt {E_m } \tilde \eta _{t,r} } \right\}}  - \frac{{E_m }}{2}\sum\limits_{r = 1}^{N_r } {\left| {\hat \alpha _{t,r} } \right|^2 }  \\
 \hat D_{m_q } \left( {m_q } \right) = \sum\limits_{r = 1}^{N_r } {{\mathop{\rm Re}\nolimits} \left\{ {\left( {\alpha _{q,r}  + \varepsilon _{q,r} } \right)^ *  \left( {\alpha _{q,r} E_m  + \sqrt {E_m } \tilde \eta _{q,r} } \right)} \right\}}  - \frac{{E_m }}{2}\sum\limits_{r = 1}^{N_r } {\left| {\hat \alpha _{q,r} } \right|^2 }  \\
 \end{array} \right.
\end{equation}
\normalsize \hrulefill \vspace*{-10pt}
\end{figure*}
\begin{figure*}[!t]
\setcounter{equation}{23}
\begin{equation} \footnotesize
\label{Eq_22}
\begin{array}{l}
 {\rm{PEP}}\left( {m_q  \to m_t } \right) \\
  = \Pr \left\{ \begin{array}{c}
 \sum\limits_{r = 1}^{N_r } {\left[ {\frac{1}{2}\left( {\alpha _{q,r} \sqrt {\bar \gamma }  + \varepsilon _{q,r} \sqrt {\bar \gamma } } \right)^ *  \left( {\alpha _{q,r} \sqrt {\bar \gamma }  + \frac{{\tilde \eta _{q,r} }}{{\sqrt {N_0 } }}} \right) + \frac{1}{2}\left( {\alpha _{q,r} \sqrt {\bar \gamma }  + \varepsilon _{q,r} \sqrt {\bar \gamma } } \right)\left( {\alpha _{q,r} \sqrt {\bar \gamma }  + \frac{{\tilde \eta _{q,r} }}{{\sqrt {N_0 } }}} \right)^ *   - \frac{1}{2}\left| {\alpha _{q,r} \sqrt {\bar \gamma }  + \varepsilon _{q,r} \sqrt {\bar \gamma }} \right|^2 } \right]}  \\
  <  \\
 \sum\limits_{r = 1}^{N_r } {\left[ {\frac{1}{2}\left( {\alpha _{t,r} \sqrt {\bar \gamma }  + \varepsilon _{t,r} \sqrt {\bar \gamma } } \right)^ *  \frac{{\tilde \eta _{t,r} }}{{\sqrt {N_0 } }} + \frac{1}{2}\left( {\alpha _{t,r} \sqrt {\bar \gamma }  + \varepsilon _{t,r} \sqrt {\bar \gamma } } \right)\frac{{\tilde \eta _{t,r}^ *  }}{{\sqrt {N_0 } }} - \frac{1}{2}\left| {\alpha _{t,r} \sqrt {\bar \gamma }  + \varepsilon _{t,r} \sqrt {\bar \gamma } } \right|^2 } \right]}  \\
 \end{array} \right\} \\
 \end{array}
\end{equation}
\normalsize \hrulefill \vspace*{-10pt}
\end{figure*}
\section{TOSD--SSK Modulation} \label{TOSD-SSK}
In this section, we focus our attention only on decision metrics and PEPs/APEPs since (\ref{Eq_7}) and (\ref{Eq_8}) are general and can be used for
TOSD--SSK modulation too.
\subsection{Decision Metrics} \label{DecisionMetrics__TOSD-SSK}
In TOSD--SSK modulation, the decision metric in (\ref{Eq_4}), can be re--written as:
\setcounter{equation}{20}
\begin{equation}
\label{Eq_19} \begin{split}
\hat D_{m_q } \left( {m_t } \right) &= \sum\limits_{r = 1}^{N_r } {{\mathop{\rm Re}\nolimits} \left\{ {\alpha _{q,r} \hat \alpha
_{t,r}^ *  E_m \delta _{t,q}  + \hat \alpha _{t,r}^ *  \sqrt {E_m } \tilde \eta _{t,r} } \right\}} \\  &- \frac{{E_m }}{2}\sum\limits_{r = 1}^{N_r
} {\left| {\hat \alpha _{t,r} } \right|^2 } \end{split}
\end{equation}
\noindent where we have taken into account that the shaping filters, $w_t \left( \cdot \right)$, are time--orthogonal to one another, and we have
defined $\tilde \eta_{t,r} = \int\nolimits_{T_m } {\eta_r \left( \xi \right)w_t^ *  \left( \xi \right)d\xi}$. In particular, for $t \ne q$ and $t = q$, the decision metric in (\ref{Eq_19}) simplifies as shown in (\ref{Eq_20}) on top of this page.
\begin{figure*}[!t]
\setcounter{equation}{31}
\begin{equation}
\label{Eq_31} {\rm{ABEP}} \le \frac{1}{{N_t \log _2 \left( {N_t } \right)}}\sum\limits_{q = 1}^{N_t } {\sum\limits_{t = 1}^{N_t } {\left[ {N_H
\left( {t,q} \right)\left( {\frac{1}{2} - \frac{1}{\pi }\int\nolimits_0^{ + \infty } { {\frac{{{\mathop{\rm Im}\nolimits} \left\{
{\Upsilon _q \left( \nu  \right)\Upsilon _t \left( { - \nu } \right)M_{\gamma _{t,q}^{\left( \Delta  \right)} \left( \nu  \right)} \left( 1
\right)} \right\}}}{\nu }} d\nu } } \right)} \right]} }
\end{equation}
\normalsize \hrulefill \vspace*{-5pt}
\end{figure*}
\begin{figure*}[!t]
\setcounter{equation}{33}
\begin{equation}
\label{Eq_33} {\rm{ABEP}} \le \frac{{N_t }}{4} - \frac{{N_t }}{{2\pi }}\int\nolimits_0^{ + \infty } { {{\mathop{\rm Im}\nolimits} \left\{
{\frac{{\Upsilon _q \left( \nu  \right)\Upsilon _t \left( { - \nu } \right)}}{\nu }\frac{1}{{\left( {1 - \Omega _0 \Delta _q \left( \nu
\right)} \right)^{N_r } \left( {1 - \Omega _0 \Delta _t \left( { - \nu } \right)} \right)^{N_r } }}} \right\}} d\nu }
\end{equation}
\normalsize \hrulefill \vspace*{-10pt}
\end{figure*}
\subsection{Computation of PEPs} \label{PEP__TOSD-SSK}
The PEPs, ${\rm{PEP}}\left( {m_q  \to m_t } \right)$, in (\ref{Eq_7}) can be computed from (\ref{Eq_3}) and (\ref{Eq_20}):
\setcounter{equation}{22}
\begin{equation}
\label{Eq_21} \begin{split} {\rm{PEP}}\left( {m_q  \to m_t } \right) &= \Pr \left\{ {\hat D_{m_q } \left( {m_q } \right) < \hat D_{m_q } \left( {m_t }
\right)} \right\} \\ &= \Pr \left\{ {\frac{{\hat D_{m_q } \left( {m_q } \right)}}{{N_0 }} < \frac{{\hat D_{m_q } \left( {m_t } \right)}}{{N_0 }}}
\right\} \end{split}
\end{equation}

By using the identity ${\mathop{\rm Re}\nolimits} \left\{ {ab^ *  } \right\} = \left( {{1 \mathord{\left/ {\vphantom {1 2}} \right.
\kern-\nulldelimiterspace} 2}} \right)ab^ *   + \left( {{1 \mathord{\left/ {\vphantom {1 2}} \right. \kern-\nulldelimiterspace} 2}} \right)a^ *
b$, which holds for every pair of complex numbers $a$ and $b$, and by explicitly showing the SNR $\bar \gamma {\rm{ = }}{{E_m } \mathord{\left/ {\vphantom {{E_m } { {N_0 }}}} \right. \kern-\nulldelimiterspace} {{N_0 }}}$ in (\ref{Eq_20}), the PEPs in (\ref{Eq_21}) simplifies as shown in (\ref{Eq_22}) on top of this page.

By introducing the RVs ($r=1,2,\ldots,N_r$): i) $X_{q,r}  = \alpha _{q,r} \sqrt {\bar \gamma }  + \varepsilon _{q,r} \sqrt {\bar \gamma }$; ii) $Y_{q,r}  = \alpha _{q,r} \sqrt {\bar \gamma }  + \left( {{{\tilde \eta _{q,r} } \mathord{\left/{\vphantom {{\tilde \eta _{q,r} } {\sqrt {N_0 } }}} \right. \kern-\nulldelimiterspace} {\sqrt {N_0 } }}} \right)$; iii) $X_{t,r}  = \alpha _{t,r} \sqrt {\bar \gamma }  + \varepsilon _{t,r} \sqrt {\bar \gamma }$; iv) $Y_{t,r}  = {{\tilde \eta _{t,r} } \mathord{\left/ {\vphantom {{\tilde \eta _{t,r} } {\sqrt {N_0 } }}} \right. \kern-\nulldelimiterspace} {\sqrt {N_0 } }}$; and:
\setcounter{equation}{24}
\begin{equation} \footnotesize
\label{Eq_24} \left\{ \begin{array}{l}
 D_q  = \sum\limits_{r = 1}^{N_r } {\left( {A\left| {X_{q,r} } \right|^2  + B\left| {Y_{q,r} } \right|^2  + CX_{q,r} Y_{q,r}^ *   + C^ *  X_{q,r}^ *  Y_{q,r} } \right)}  \\
 D_t  = \sum\limits_{r = 1}^{N_r } {\left( {A\left| {X_{t,r} } \right|^2  + B\left| {Y_{t,r} } \right|^2  + CX_{t,r} Y_{t,r}^ *   + C^ *  X_{t,r}^ *  Y_{t,r} } \right)}  \\
 \end{array} \right.
\end{equation}
\noindent the PEP in (\ref{Eq_22}) can be re--written (with $A = - {1 \mathord{\left/ {\vphantom {1 2}} \right. \kern-\nulldelimiterspace} 2}$, $B = 0$, and $C = {1 \mathord{\left/ {\vphantom {1 2}} \right. \kern-\nulldelimiterspace} 2}$) as:
\setcounter{equation}{25}
\begin{equation}
\label{Eq_25} \begin{split} {\rm{PEP}}\left( {m_q  \to m_t } \right) &= \Pr \left\{ {D_q  < D_t } \right\} \\ &= \Pr \left\{ {D_{t,q}  = D_q  - D_t  < 0} \right\} \end{split}
\end{equation}

Similar to Section \ref{PEP__SSK}, from \cite[Sec. III]{MDR_QF}, we can readily conclude that both $D_q$ and $D_t$ in (\ref{Eq_24}) are
quadratic--forms in conditional complex Gaussian RVs. Furthermore, we note that $D_q$ and $D_t$ are, when conditioning upon the fading channel
gains, statistically independent, as AWGN and channel estimation errors are independent from one another if $t \ne q$ and across the
$N_r$ receive--antennas. We emphasize that to compute the ABEP in (\ref{Eq_7}) we are interested only in the cases where $t \ne q$, as $N_H
\left( {t,q} \right) = 0$ if $t=q$.

From (\ref{Eq_25}), the PEPs can be still computed by using the Gil--Pelaez inversion theorem \cite{GilPelaez}:
\setcounter{equation}{26}
\begin{equation}
\label{Eq_26} \begin{split} & {\rm{PEP}}\left( {m_q  \to m_t } \right) = \frac{1}{2} - \frac{1}{\pi }\int\nolimits_0^{ + \infty } { {\frac{{{\mathop{\rm
Im}\nolimits} \left\{ {\Psi _{D_{t,q} } \left( {\left. \nu  \right|{\boldsymbol{\alpha }}_{t,q} } \right)} \right\}}}{\nu }d\nu } } \\ & \hspace{0.5cm}
\mathop  = \limits^{\left( a \right)} \frac{1}{2} - \frac{1}{\pi }\int\nolimits_0^{ + \infty } { {\frac{{{\mathop{\rm Im}\nolimits}
\left\{ {\Psi _{D_q } \left( {\left. \nu  \right|{\boldsymbol{\alpha }}_q } \right)\Psi _{D_t } \left( {\left. { - \nu }
\right|{\boldsymbol{\alpha }}_t } \right)} \right\}}}{\nu }d\nu } } \end{split}
\end{equation}
\noindent where $\Psi _{D_{t,q}} \left( {\left.  \cdot  \right|{\boldsymbol{\alpha}}_{t,q}} \right)$ is the CF of RV $D_{t,q}$ when
conditioning upon the fading gains ${\boldsymbol{\alpha}}_{t,q} = \left\{ {\alpha _{t,r} ,\alpha _{q,r} } \right\}_{r = 1}^{N_r }$, and
${\Psi _{D_q } \left( {\left. \cdot  \right|{\boldsymbol{\alpha }}_q } \right)}$ and ${\Psi _{D_t } \left( {\left. { \cdot }
\right|{\boldsymbol{\alpha }}_t } \right)}$ are the CFs of RVs $D_q$ and $D_t$ when conditioning upon the fading gains ${\boldsymbol{\alpha
}}_q  = \left\{ {\alpha _{q,r} } \right\}_{r = 1}^{N_r }$ and ${\boldsymbol{\alpha }}_t  = \left\{ {\alpha _{t,r} } \right\}_{r = 1}^{N_r }$,
respectively. Furthermore, $\mathop  = \limits^{\left( a \right)}$ comes from the independence of the conditional RVs $D_q$ and
$D_t$, and the definition of CF, \emph{i.e.}, $\Psi _{D_{t,q} } \left( {\left. \nu  \right|{\boldsymbol{\alpha }}_{t,q} } \right) =
{\rm{E}}_{\eta {\rm{,}}\varepsilon } \left\{ {\exp \left( {j\nu D_{t,q} } \right)} \right\} = {\rm{E}}_{\eta {\rm{,}}\varepsilon } \left\{ {\exp \left( {j\nu D_q } \right)\exp \left( { - j\nu D_t } \right)} \right\} = \Psi _{D_q } \left( {\left. \nu  \right|{\boldsymbol{\alpha }}_q } \right)\Psi _{D_t }
\left( {\left. { - \nu } \right|{\boldsymbol{\alpha }}_t } \right)$. We emphasize that ${\rm{E}}_{\eta {\rm{,}}\varepsilon } \left\{  \cdot
\right\}$ is the expectation operator computed over AWGN and channel estimation errors, as we are conditioning upon the channel gains.

The last step is to compute the CFs in (\ref{Eq_26}), which can be obtained from \cite[Eq. (2) and Eq.(3)]{MDR_QF} by using the theory of quadratic--forms in conditional complex Gaussian RVs, as:
\setcounter{equation}{27}
\begin{equation}
\label{Eq_27} \left\{ \begin{array}{l}
 \Psi _{D_q } \left( {\left. \nu  \right|{\boldsymbol{\alpha }}_q } \right) = \Upsilon _q \left( \nu  \right)\exp \left\{ {\Delta _q \left( \nu  \right)\gamma _q } \right\} \\
 \Psi _{D_t } \left( {\left. \nu  \right|{\boldsymbol{\alpha }}_t } \right) = \Upsilon _t \left( \nu  \right)\exp \left\{ {\Delta _t \left( \nu  \right)\gamma _t } \right\} \\
 \end{array} \right.
\end{equation}
\noindent where we have defined: i) $v_a  = \sqrt {\left( {{1 \mathord{\left/ {\vphantom {1 4}} \right. \kern-\nulldelimiterspace} 4}} \right) + N_p r_{pm} }  + \left( {{1 \mathord{\left/ {\vphantom {1 2}} \right. \kern-\nulldelimiterspace} 2}} \right)$; ii) $v_b  = \sqrt {\left( {{1 \mathord{\left/ {\vphantom {1 4}} \right. \kern-\nulldelimiterspace} 4}} \right) + N_p r_{pm} }  - \left( {{1 \mathord{\left/ {\vphantom {1 2}} \right. \kern-\nulldelimiterspace} 2}} \right)$; iii) $\gamma _q  = \gamma \left( {{\boldsymbol{\alpha }}_q } \right) = \sum\nolimits_{r = 1}^{N_r } {\left| {\alpha _{q,r} } \right|^2 }$; iv) $\gamma _t  = \gamma \left( {{\boldsymbol{\alpha }}_t } \right) = \sum\nolimits_{r = 1}^{N_r } {\left| {\alpha _{t,r} } \right|^2 }$; v) $g_a^{\left( q \right)}  = \left( {{1 \mathord{\left/ {\vphantom {1 2}} \right. \kern-\nulldelimiterspace} 2}} \right)\bar \gamma \left[ {1 + \left( {N_p r_{pm} } \right)^{ - 1} } \right]$; vi) $g_b^{\left( q \right)}  = g_a^{\left( t \right)}  =  - g_b^{\left( t \right)}  = \left( {{1 \mathord{\left/ {\vphantom {1 2}} \right. \kern-\nulldelimiterspace} 2}} \right)\bar \gamma$; and:
\setcounter{equation}{28}
\begin{equation} \footnotesize
\label{Eq_28}
\left\{ \begin{array}{l}
 \Delta _q \left( \nu  \right) = v_a v_b \left( { - \nu ^2 g_a^{\left( q \right)}  + j\nu g_b^{\left( q \right)} } \right)\left( {\nu  + jv_a } \right)^{ - 1} \left( {\nu  - jv_b } \right)^{ - 1} \\
 \Delta _t \left( \nu  \right) = v_a v_b \left( { - \nu ^2 g_a^{\left( t \right)}  + j\nu g_b^{\left( t \right)} } \right)\left( {\nu  + jv_a } \right)^{ - 1} \left( {\nu  - jv_b } \right)^{ - 1} \\
 \Upsilon _q \left( \nu  \right) = \Upsilon _t \left( \nu  \right) = \left( {v_a v_b } \right)^{N_r } \left( {\nu  + jv_a } \right)^{ - N_r } \left( {\nu  - jv_b } \right)^{ - N_r } \\
 \end{array} \right.
\end{equation}
\subsection{Computation of APEPs} \label{APEP__TOSD-SSK}
The APEP can be computed from (\ref{Eq_26}) and (\ref{Eq_27}) by still using the Gil--Pelaez inversion theorem \cite{GilPelaez}:
\setcounter{equation}{29}
\begin{equation}
\label{Eq_29} \begin{split} {\rm{APEP}}\left( {m_q  \to m_t } \right) &= {\rm{E}}\left\{ {{\rm{PEP}}\left( {m_q  \to m_t } \right)} \right\} \\ &= \frac{1}{2} -
\frac{1}{\pi }\int\nolimits_0^{ + \infty } { {\frac{{{\mathop{\rm Im}\nolimits} \left\{ {\Psi _{D_{t,q} } \left( \nu  \right)}
\right\}}}{\nu }d\nu } } \end{split}
\end{equation}
\noindent where $\Psi _{D_{t,q} } \left( \nu  \right) = {\rm{E}}\left\{ {\Psi _{D_q } \left( {\left. \nu  \right|{\boldsymbol{\alpha }}_q }
\right)\Psi _{D_t } \left( {\left. { - \nu } \right|{\boldsymbol{\alpha }}_t } \right)} \right\}$, which, for generic fading channels, is:
\setcounter{equation}{30}
\begin{equation}
\label{Eq_30}
\begin{split}
 & \Psi _{D_{t,q} } \left( \nu  \right) = {\rm{E}}\left\{ {\Psi _{D_q } \left( {\left. \nu  \right|{\boldsymbol{\alpha }}_q } \right)\Psi _{D_t } \left( {\left. { - \nu } \right|{\boldsymbol{\alpha }}_t } \right)} \right\} \\
 & \hspace{0.5cm} = {\rm{E}}\left\{ {\Upsilon _q \left( \nu  \right)\exp \left\{ {\Delta _q \left( \nu  \right)\gamma _q } \right\}\Upsilon _t \left( { - \nu } \right)\exp \left\{ {\Delta _t \left( { - \nu } \right)\gamma _t } \right\}} \right\} \\
 & \hspace{0.5cm} = \Upsilon _q \left( \nu  \right)\Upsilon _t \left( { - \nu } \right){\rm{E}}\left\{ {\exp \left\{ {\Delta _q \left( \nu  \right)\gamma _q  + \Delta _t \left( { - \nu } \right)\gamma _t } \right\}} \right\} \\
 & \hspace{0.5cm} = \Upsilon _q \left( \nu  \right)\Upsilon _t \left( { - \nu } \right)M_{\gamma _{t,q}^{\left( \Delta  \right)} \left( \nu  \right)} \left( 1 \right)
 \end{split}
\end{equation}
\noindent and $M_{\gamma _{t,q}^{\left( \Delta  \right)} \left( \nu  \right)} \left( s \right) = {\rm{E}}\left\{ {\exp \left( {s\gamma
_{t,q}^{\left( \Delta  \right)} \left( \nu  \right)} \right)} \right\}$ is the MGF of RV $\gamma _{t,q}^{\left( \Delta  \right)} \left( \nu
\right) = \Delta _q \left( \nu  \right)\gamma _q  + \Delta _t \left( { - \nu } \right)\gamma _t$.

In conclusion, the ABEP of TOSD--SSK modulation over arbitrary fading channels and with practical channel estimates can be computed in
closed--form from (\ref{Eq_7}), (\ref{Eq_29}), and (\ref{Eq_30}) as shown in (\ref{Eq_31}) on top of this page.

Similar to SSK modulation, (\ref{Eq_31}) is general and useful for every MIMO setups. To be computed, a
closed--form expression of the MGF of RV $\gamma _{t,q}^{\left( \Delta  \right)} \left( \nu  \right) = \Delta _q \left( \nu  \right)\gamma _q +
\Delta _t \left( { - \nu } \right)\gamma _t$, which is given by the linear combination of the power--sum of generically correlated and
distributed channel gains, is needed. This MGF is available for various fading channel models in \cite{Simon}, or, \emph{e.g.}, it can be readily computed by
exploiting the Moschopoulos method for arbitrarily correlated and distributed Rician fading channels, as described in \cite{MDR_TOSD_TCOM}. In
particular, if all the channel gains are independent, but not necessarily identically distributed, the MGF $M_{\gamma _{t,q}^{\left( \Delta
\right)} \left( \nu \right)} \left( \cdot \right)$ reduces to:
\setcounter{equation}{32}
\begin{equation}
\label{Eq_32}
\begin{split}
 & M_{\gamma _{t,q}^{\left( \Delta  \right)} \left( \nu  \right)} \left( s \right) = {\rm{E}}\left\{ {\exp \left( {s\gamma _{t,q}^{\left( \Delta  \right)} \left( \nu  \right)} \right)} \right\} \\
 & \hspace{0.33cm} = {\rm{E}}\left\{ {\exp \left\{ {s\Delta _q \left( \nu  \right)\gamma _q } \right\}} \right\}{\rm{E}}\left\{ {\exp \left\{ {s\Delta _t \left( { - \nu } \right)\gamma _t } \right\}} \right\} \\
 & \hspace{0.33cm} = \left[ {\prod\limits_{r = 1}^{N_r } {M_{\left| {\alpha _{q,r} } \right|^2 } \left( {s\Delta _q \left( \nu  \right)} \right)} } \right] \cdot \left[ {\prod\limits_{r = 1}^{N_r } {M_{\left| {\alpha _{t,r} } \right|^2 } \left( {s\Delta _t \left( { - \nu } \right)} \right)} } \right] \\
 \end{split}
\end{equation}
\noindent where the MGFs $M_{\left| {\alpha _{t,r} } \right|^2 } \left(  \cdot  \right)$ and $M_{\left| {\alpha _{q,r} } \right|^2 } \left(
\cdot  \right)$ are available in closed--form in \cite{Simon} for almost all fading channel models of interest in wireless communications. For
example, if the channel gains are i.i.d. Rayleigh distributed the ABEP in (\ref{Eq_31}) reduces, from (\ref{Eq_8}), to (\ref{Eq_33}) shown on top of this page.

Finally, similar to Section \ref{APEP__SSK}, we note that: i) the integrand function in (\ref{Eq_31}) is, for typical MGFs used in
communication problems, well--behaved when $\nu \to 0$; ii) since the ABEP in, \emph{e.g.}, (\ref{Eq_32}) and (\ref{Eq_33}) is given by the
products of $2N_r$ MGFs, we conclude from \cite{MDR_TOSD_TCOM} and \cite{Giannakis} that the diversity order of the system is $2N_r$, which is
the same as the P--CSI scenario \cite{MDR_TOSD_TCOM}; and iii) (\ref{Eq_31}) reduces to the P--CSI lower--bound if $N_p r_{pm} \to + \infty$.
\begin{table*}
\renewcommand{\arraystretch}{1.2}
\caption{Bandwidth of various time--limited shaping filters. Time and frequency responses of rectangular, half--sine, and raised--cosine
shaping filters are available in the captions of Fig. \ref{Fig_PulsesFreq_Reference}. The
shaping filters ${w_{t } \left( \cdot \right)}$ are given in (\ref{Eq_App1}). Let $P\left( \omega  \right) = \left( 1/ \sqrt {{\rm{2}}\pi
}\right)\int\nolimits_{ - \infty }^{ + \infty } {p\left( \xi \right)\exp \left( { - j\omega \xi } \right)d\xi }$ be the Fourier transform of a
generic shaping filter with time response $p\left(  \cdot  \right)$. Then: i) the Fractional Power Containment Bandwidth (FPCB) is defined as
${\rm{FPCB}}_{{\rm{X}}_{\rm{\% }} }  = \mathop {\min }\limits_{B \in \left[ {0, + \infty } \right)} \left\{ {B|\frac{{\int\nolimits_0^B {\left|
{P\left( \omega  \right)} \right|^2 d\omega } }}{{\int\nolimits_0^{ + \infty } {\left| {P\left( \omega \right)} \right|^2 d\omega } }} >
{\rm{X}}_{\rm{\% }} } \right\}$ \cite[p. 15]{Amoroso_BW}, which is the bandwidth $B$ where ${{\rm{X}}_{\rm{\% }} }$ percent of the energy is
contained; and ii) the Bounded Power Spectral Density Bandwidth (BPSDB) is defined as ${\rm{BPSDB}}_{{\rm{TH}}_{{\rm{dB}}} } = \mathop {\min
}\limits_{B \in \left[ {0, + \infty } \right)} \left\{ {B|\log _{10} \left( {\left| {P\left( \omega  \right)} \right|^2 } \right) < \log _{10}
\left( {\left| {P\left( {\omega _{{\rm{peak}}} } \right)} \right|^2 } \right) - {\rm{TH}}_{{\rm{dB}}} ,\;\forall \omega  > B\;} \right\}$
\cite[p. 18]{Amoroso_BW}, which is the bandwidth $B$ beyond which the spectral density is ${{\rm{TH}}_{{\rm{dB}}} }$ below its peak (maximum
value), \emph{i.e.}, ${\left| {P\left( {\omega _{{\rm{peak}}} } \right)} \right|^2 }$.} \label{Tab_0}
\begin{center}
\begin{tabular}{|c||c|c|c|c|}
\hline
\multicolumn{5}{|c|} {Fractional Power Containment Bandwidth ($B/(2\pi)$ kHz)} \\
\hline
${{\rm{X}}_{\rm{\% }} }$ & Rectangular & Half-Sine & Raised-Cosine & ${w_{t } \left( \cdot  \right)}$ \\
\hline
99\% & 7.61 & 1.18 & 1.41 & 4.97 \\
\hline
99.995\% & $>$30 & 6.98 & 3.29 & 6.46 \\
\hline
99.9999\% & $>$30 & 22.14 & 6.64 & 7.31 \\
\hline
99.99999\% & $>$30 & 29.96 & 10.57 & 7.76 \\
\hline \hline
\multicolumn{5}{|c|} {Bounded Power Spectral Density Bandwidth ($B/(2\pi)$ kHz)} \\
\hline
${{\rm{TH}}_{{\rm{dB}}} }$ & Rectangular & Half-Sine & Raised-Cosine & ${w_{t } \left( \cdot  \right)}$ \\
\hline
3dB & 9.59 & 2.28 & 1.85 & 6.35 \\
\hline
5dB & $>$30 & 8.18 & 4.62 & 7.40 \\
\hline
6dB & $>$30 & 15.13 & 6.64 & 7.85 \\
\hline
7dB & $>$30 & 28.03 & 9.65 & 8.27 \\
\hline
10dB & $>$30 & $>$30 & $>$30 & 9.39 \\
\hline
\end{tabular}
\end{center}
\end{table*}
\section{Bandwidth Efficiency of Orthogonal Shaping Filters Design} \label{BandwidthPulses}
In Section \ref{TOSD-SSK}, we have shown that TOSD--SSK modulation provides, even in the presence of channel estimation errors and with a
single active antenna at the transmitter, a diversity order that is equal to $2N_r$. This is achieved by using time--orthogonal shaping filters
at the transmitter, which is an additional design constraint that might not be required by SSK modulation and conventional single-- and
multiple--antenna systems. Thus, for a fair comparison among the various modulation schemes, it is important to assess whether the
time--orthogonal constraint affects the overall bandwidth efficiency of the communication system. More specifically, this section is aimed at
understanding whether a larger transmission bandwidth is required for the transmission of the same number of bits in a given signaling
time--interval $T_m$, \emph{i.e.}, for a given bit/symbol or bpcu requirement. To shed light on this matter, in this
section we analyze the bandwidth occupancy of commonly used shaping filters and, as an illustrative example, a family of recently proposed
spectrally--efficient orthogonal shaping filters. More specifically: i) as far state--of--the--art shaping filters are concerned, we consider
well--known time--limited rectangular, half--sine, and raised--cosine prototypes \cite[Sec. III--B]{Hanzo_BW}; on the other hand, ii) as far as
time--orthogonal shaping filters are concerned, we consider waveforms built upon linear combinations of Hermite polynomials
\cite{MDR_TOSD_TCOM}, \cite{TCOM_PSM}. The analytical expressions of time and frequency responses of these letter filters are available in Appendix \ref{App} for $N_t=4$.

\begin{figure}[!t]
\centering
\includegraphics [width=\columnwidth] {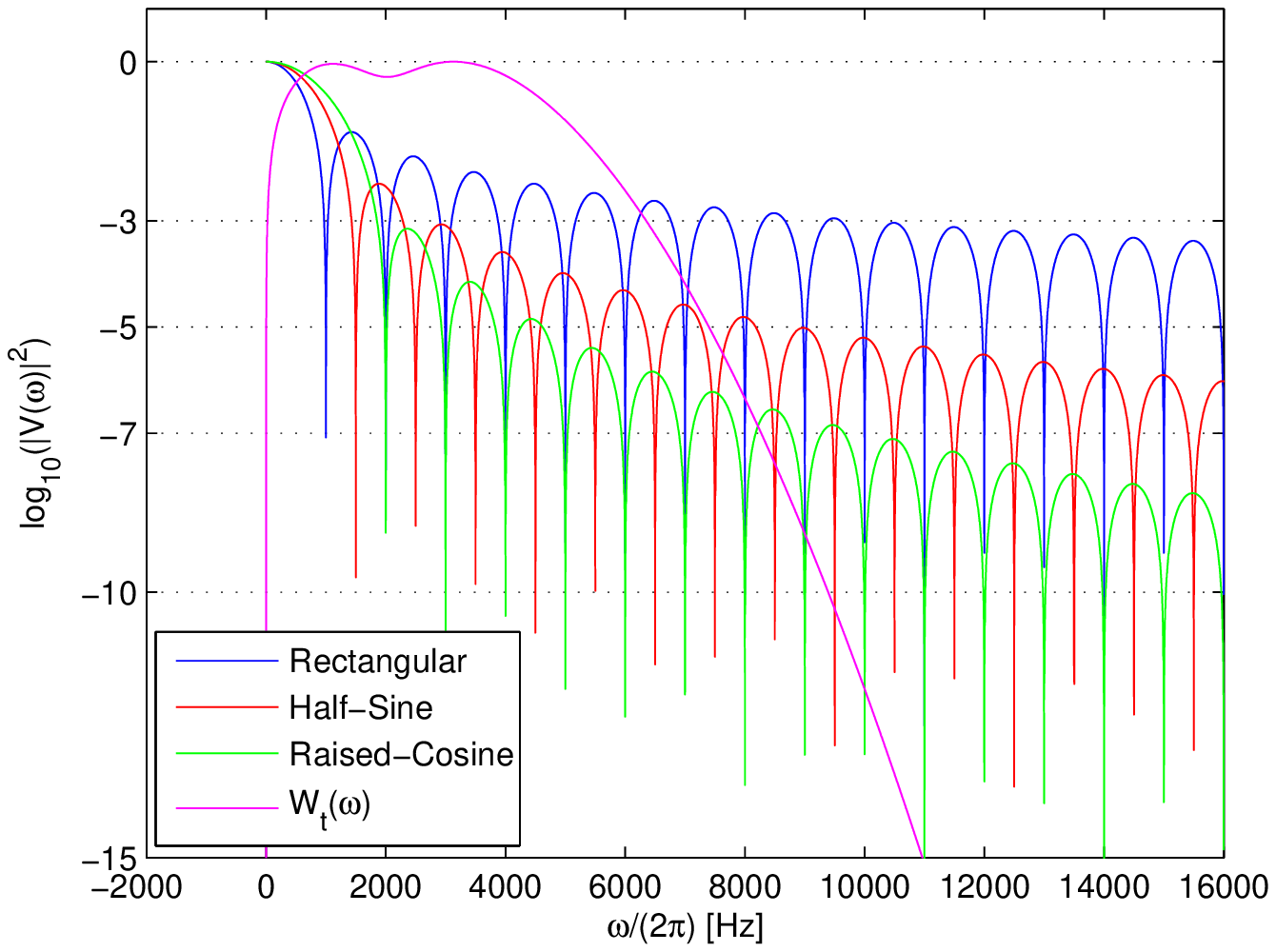}
\caption{Examples of time--limited shaping filters commonly used in the literature (frequency responses). The time responses are as follows. i) Rectangular pulse: $v\left( \xi \right) = p_{T_0 } \left( \xi  \right)$, where $p_{T_0 } \left( \xi  \right) = 1$ if $ - T_0 /2 \le \xi \le T_0 /2$ and $p_{T_0 } \left( \xi \right) = 1$ elsewhere. ii) Half--sine pulse: $v\left( \xi  \right) = \sqrt 2 \sin \left[ {\pi \left( {\xi  + 0.5T_0 } \right)/T_0 } \right]p_{T_0 } \left( \xi  \right)$. iii) Raised--cosine pulse: $v\left( \xi \right) = \sqrt {2/3} \left\{ {1 - \cos \left[ {2\pi \left( {\xi + 0.5T_0 } \right)/T_0 } \right]} \right\}p_{T_0 } \left( \xi  \right)$. iv) The orthogonal shaping filters for $N_t=4$ are given in (\ref{Eq_App1}) in Appendix \ref{App} with $t_0=10^{-4}$. $T_0=10^{-3}$ is the time duration of the filters. The frequency response is defined as $V\left( \omega  \right) = \left( 1/{\sqrt {2\pi } } \right)\int\nolimits_{ - \infty }^{ + \infty } {v\left( \xi \right)\exp \left( { - j\omega \xi } \right)d\xi }$. They are as follows. i) Rectangular pulse: $\left| {V\left( {2\pi \omega } \right)} \right| = \kappa _r {\rm{sinc}}\left( {\omega T_0 } \right)$, where ${\rm{sinc}}\left( x \right) = 1$ if $x=0$ and ${\rm{sinc}}\left( x \right) = \sin \left( {\pi x} \right)/\pi x$ if $x \ne 0$. ii) Half--sine pulse: $\left| {V\left( {2\pi \omega }
\right)} \right| = \kappa _{hs} \left[ {\cos \left( {\pi \omega T_0 } \right)/\left( {1 - 4\omega ^2 T_0^2 } \right)} \right]$. iii)
Raised--cosine pulse: $\left| {V\left( {2\pi \omega } \right)} \right| = \kappa _{rc} \left[ {{\rm{sinc}}\left( {\omega T_0 } \right) + \left(
{1/2} \right){\rm{sinc}}\left( {\omega T_0  - 1} \right) + \left( {1/2} \right){\rm{sinc}}\left( {\omega T_0  + 1} \right)} \right]$. iv) The orthogonal shaping filters for $N_t=4$ are given in (\ref{Eq_App3}) in Appendix \ref{App} with $t_0=10^{-4}$. $\kappa_r$, $\kappa_{hs}$, and $\kappa_{rc}$ are constant factors that are not relevant for our analysis.} \label{Fig_PulsesFreq_Reference}
\end{figure}

Three important comments are worth being made about the shaping filters that are considered in our comparative study: 1) we limit our study to
considering time--limited shaping filters, which are simpler to be implemented than bandwidth--limited filters \cite{Hanzo_BW}, \cite{Kang_BW}.
This choice allows us to perform a fair comparison among SSK modulation and conventional modulation schemes. In fact, an important benefit of
SSK and TOSD--SSK modulations is to take advantage of multiple--antenna technology with a single Radio Frequency (RF) front end at the
transmitter \cite{Haas_TVT}, \cite{Ghrayeb_TWC}, which is a research challenge that is currently stimulating the development of novel MIMO
concepts based, \emph{e.g.}, on parasitic antenna architectures \cite{Papadias_Aug2008}--\cite{Papadias_Feb2011}. A recent survey on single--RF MIMO design is available in \cite{SingleRF_Dec2011}. In order to use a single--RF
chain, SSK and TOSD--SSK modulations need shaping filters that are time--limited and have a duration that is equal to the signaling
time--interval $T_m$. In fact, as remarked in \cite[Section II--D]{Ghrayeb_TWC}, the adoption of shaping filters that are not time--limited
would require a number of RF chains that is equal to the number of signaling time--intervals $T_m$ where the filter has a non--zero time
response (\emph{i.e.}, the time--duration of the filter). Thus, bandwidth--limited shaping filters \cite{Kang_BW} would require multiple RF
chains; 2) even though the orthogonal shaping filters considered in the present paper and summarized in Appendix \ref{App} are obtained by using the algorithm proposed in \cite{TCOM_PSM}, which was introduced for Ultra Wide Band (UWB) systems, time--duration and bandwidth can be adequately scaled for narrow--band communication systems. For example, Fig. \ref{Fig_PulsesFreq_Reference} is representative of a narrow--band system with
pulses having a practical time--duration of milliseconds and a practical bandwidth of kilohertz. Thus, neither UWB nor Spread Spectrum (SS)
systems with orthogonal spreading codes are needed for space modulation; and 3) the method proposed in
\cite{TCOM_PSM} for the design of orthogonal shaping filters guarantees that all the waveforms have the same time--duration and (practical)
bandwidth. Thus, unlike conventional Hermite polynomials, time--orthogonality is guaranteed without bandwidth expansion. Let us emphasize that
other methods are available in the literature to generate time--limited and time--orthogonal shaping filters. Two examples, which allow us to
jointly tuning time--duration and bandwidth and to guaranteeing low out--of--band interference, are given in \cite{Parr_BW} and
\cite{Giannakis_BW}.

Let us now compare the bandwidth efficiency of the orthogonal shaping filters available in Appendix \ref{App} with state--of--the--art shaping filters. A qualitative and quantitative comparisons are shown in Fig. \ref{Fig_PulsesFreq_Reference} and in Table \ref{Tab_0}, respectively, by using two commonly adopted definitions of bandwidth \cite{Amoroso_BW}: i) the Fractional Power Containment Bandwidth (FPCB) \cite[p. 15]{Amoroso_BW}; and ii) the Bounded Power Spectral Density Bandwidth (BPSDB) \cite[p. 18]{Amoroso_BW}. The formal definition of these two concepts of bandwidth is given in the caption of Table
\ref{Tab_0}. By carefully analyzing both Table \ref{Tab_0} and Fig. \ref{Fig_PulsesFreq_Reference}, we notice that the bandwidth efficiency of
the different shaping filters depend on how stringent the criterion to define the bandwidth is. In particular, if the percentage of energy that
is required to be contained in the bandwidth (FPCB) is 99\%, then the best shaping filter to use is the half--sine. On the other hand, if, to
reduce the interference produced in adjacent transmission bands, the requirement moves from 99\% to 99.99999\%, then the best shaping filters
to us are those given in Appendix \ref{App}. A similar comment applies when the BPSDB definition of bandwidth is used, but the best shaping
filters are the raised--cosine (less stringent requirement) and the orthogonal filters in Appendix \ref{App} (more stringent requirement). A
similar trade--off has been shown in \cite{Hanzo_BW} and \cite{Amoroso_BW} for conventional modulation schemes and shaping filters.

In other words, the shaping filters in Appendix \ref{App} are designed to have a very flat spectrum in the transmission band to improve the
energy efficiency, as well as a very fast roll--off to reduce interference and enhance coexistence capabilities. This is especially useful to increase
the system efficiency since current standards require the transmitted spectrum to occupy a well--defined spectral mask, \emph{e.g.}, for
Wireless Local Area Networks (WLAN) and UWB wireless systems. For this reason, the shaping filters in Appendix \ref{App} have a very good
energy containment and bounded energy spectrum. Finally, we emphasize that the shaping filters given in Appendix \ref{App} are just an example
of time--orthogonal filters that can be obtained with state--of--the--art signal processing algorithms \cite{Parr_BW}, \cite{Giannakis_BW}, as
well as that the waveforms compared in Table \ref{Tab_0} have the same time--duration $T_m$, and, thus, they provide the same signaling rate
$1/T_m$.

For illustrative purposes, in this paper we choose the shaping filters with the main objective to limit, as much as possible, out--of--band
interference in order to enhance the coexistence capabilities of our communication system, and to reduce interference in adjacent transmission
bands. Thus, our criterion is based on choosing filters which, for the same time--duration, have a stringent energy containment or bounded
energy spectrum. For example, we assume either ${{\rm{X}}_{\rm{\% }} } > 99.9999\%$ or ${{\rm{TH}}_{{\rm{dB}}} } > 6{{\rm{dB}}}$ in Table
\ref{Tab_0}. With these assumptions, the orthogonal shaping filters given in Appendix \ref{App} are the best choice, and are chosen to obtain
the simulation results in Section \ref{Results}. For applications where less stringent coexistence capabilities might be required, the shaping
filters given in Appendix \ref{App} might not be the best choice, as they would require a larger bandwidth. In that case, by using the
algorithms in \cite{TCOM_PSM}, \cite{Parr_BW}, \cite{Giannakis_BW}, and references therein, new orthogonal pulses could be generated with the
required time--duration and (practical) bandwidth.
\section{Numerical and Simulation Results} \label{Results}
In this section, we show some numerical examples in order to: i) study the performance of SSK and TOSD--SSK modulations in the presence of
channel estimation errors; ii) compare the achievable performance with single--antenna and Alamouti schemes; and iii) assess the
accuracy of our analytical derivation. For illustrative purposes, i.i.d Rayleigh fading channels are considered in all the analyzed scenarios. The interested reader might find in \cite{MDR_TCOM2010}, \cite{MDR_TOSD_TCOM}, and \cite{MDR_TCOM_PCSI} numerical examples about the performance of SSK and TOSD--SSK modulations for different wireless channels. Single--antenna and Alamouti schemes are chosen as state--of--the--art transmission technologies for
performance comparison because they have the same diversity order and the same decoding complexity as SSK and TOSD--SSK modulations, respectively. The interested reader might find in \cite[Fig. 2]{MDR_TOSDSM_ICC2011} the comparison with transmit--diversity Space--Time--Block--Codes (STBCs) for MIMO systems with more than two antennas at the transmitter, and in \cite[Fig. 8]{MDR_TVT} the comparison with spatial multiplexing MIMO systems with multi--user detection. In these latter cases, both STBCs and spatial multiplexing MIMO have higher decoding complexity and worse performance than space modulation.

The simulation setup used in our study is as follows: i) we consider i.i.d Rayleigh fading with unit--power over all the wireless links. The
related analytical framework is available, by setting $\Omega_0 = 1$, in (\ref{Eq_18}) and (\ref{Eq_33}) for SSK and TOSD--SSK modulations,
respectively; ii) $r_{pm} = 1$ for all the analyzed scenarios; iii) the bpcu of SSK and TOSD--SSK modulations are equal to $R = \log _2 \left(
{N_t } \right)$; iv) as far single--antenna and Alamouti schemes are concerned, we consider QAM with constellation size $M$ and bpcu equal to
$R = \log _2 \left( {M } \right)$; v) as mentioned in Section \ref{BandwidthPulses}, the shaping filters are obtained from \cite{TCOM_PSM}. For
example, when $N_t=4$, $w_{t_1 } \left(  \cdot  \right)$ in Appendix \ref{App} is used for SSK modulation, single--antenna, and Alamouti
schemes, while the set of four orthogonal filters in (\ref{Eq_App1}) is used for TOSD--SSK modulation. Furthermore, for a fair comparison among the
modulation schemes, the same spectral efficiency (measured in bpcu) is considered; vi) the ABEP of the P--CSI scenario is computed by assuming
an infinite number of pilot pulses; vii) the ABEP of single--antenna and Alamouti schemes is obtained through Monte Carlo simulations only, but
to check that our simulator is well--tuned the numerical results are compared, for the P--CSI scenario, to the ABEP predicted by the
union--bound recently developed in \cite{MDR_TVT} for single-- and multi--user systems; viii) as far as single--antenna schemes are concerned,
$E_m$ is the average energy transmitted for each information symbol; and ix) as far as the Alamouti scheme is concerned, $E_m$ is the average
energy transmitted for each information symbol from the two active transmit--antennas, \emph{i.e.}, $E_m$ is equally split between the two
antennas.

\begin{figure}[!t]
\centering
\includegraphics [width=\columnwidth] {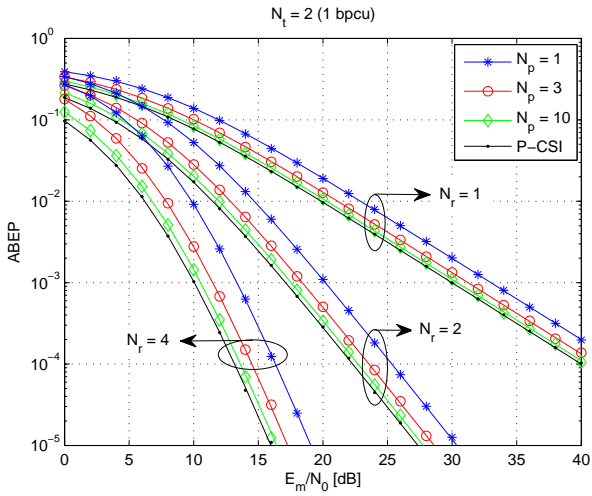}
\caption{ABEP of SSK modulation against $E_m/N_0$ for: i) $N_t = 2$ (1 bpcu); ii) $N_r = \left\{ {1,2,4} \right\}$; iii) $N_p  = \left\{
{1,3,10} \right\}$; and iv) P--CSI denotes the ABEP with no channel estimation errors. Solid lines show the analytical model and markers show Monte
Carlo simulations.} \label{Fig_SSK1bps}
\end{figure}
\begin{figure}[!t]
\centering
\includegraphics [width=\columnwidth] {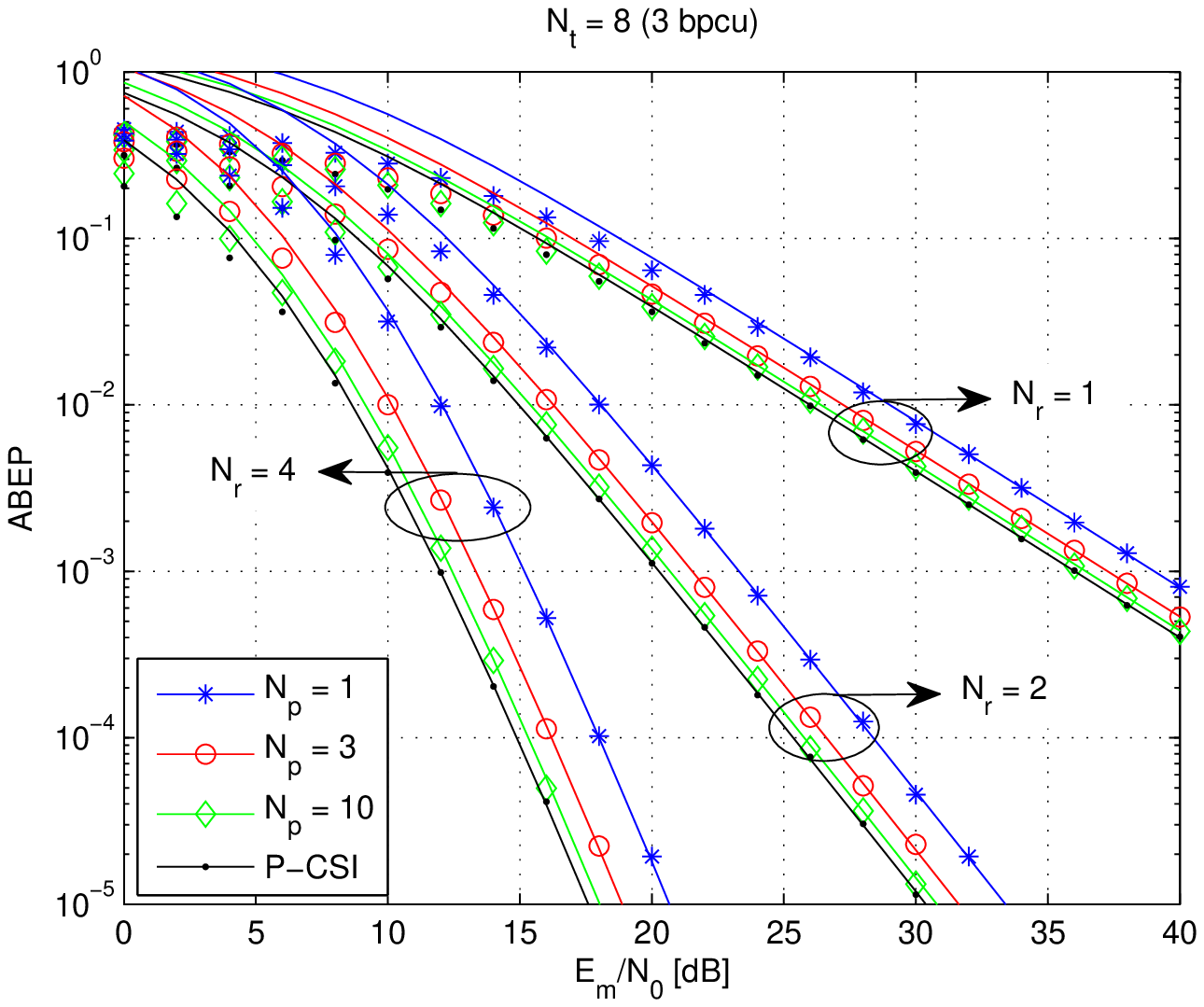}
\caption{ABEP of SSK modulation against $E_m/N_0$ for: i) $N_t = 8$ (3 bpcu); ii) $N_r = \left\{ {1,2,4} \right\}$; iii) $N_p  = \left\{
{1,3,10} \right\}$; and iv) P--CSI denotes the ABEP with no channel estimation errors. Solid lines show the analytical model and markers show Monte
Carlo simulations.} \label{Fig_SSK3bps}
\end{figure}
\begin{figure}[!t]
\centering
\includegraphics [width=\columnwidth] {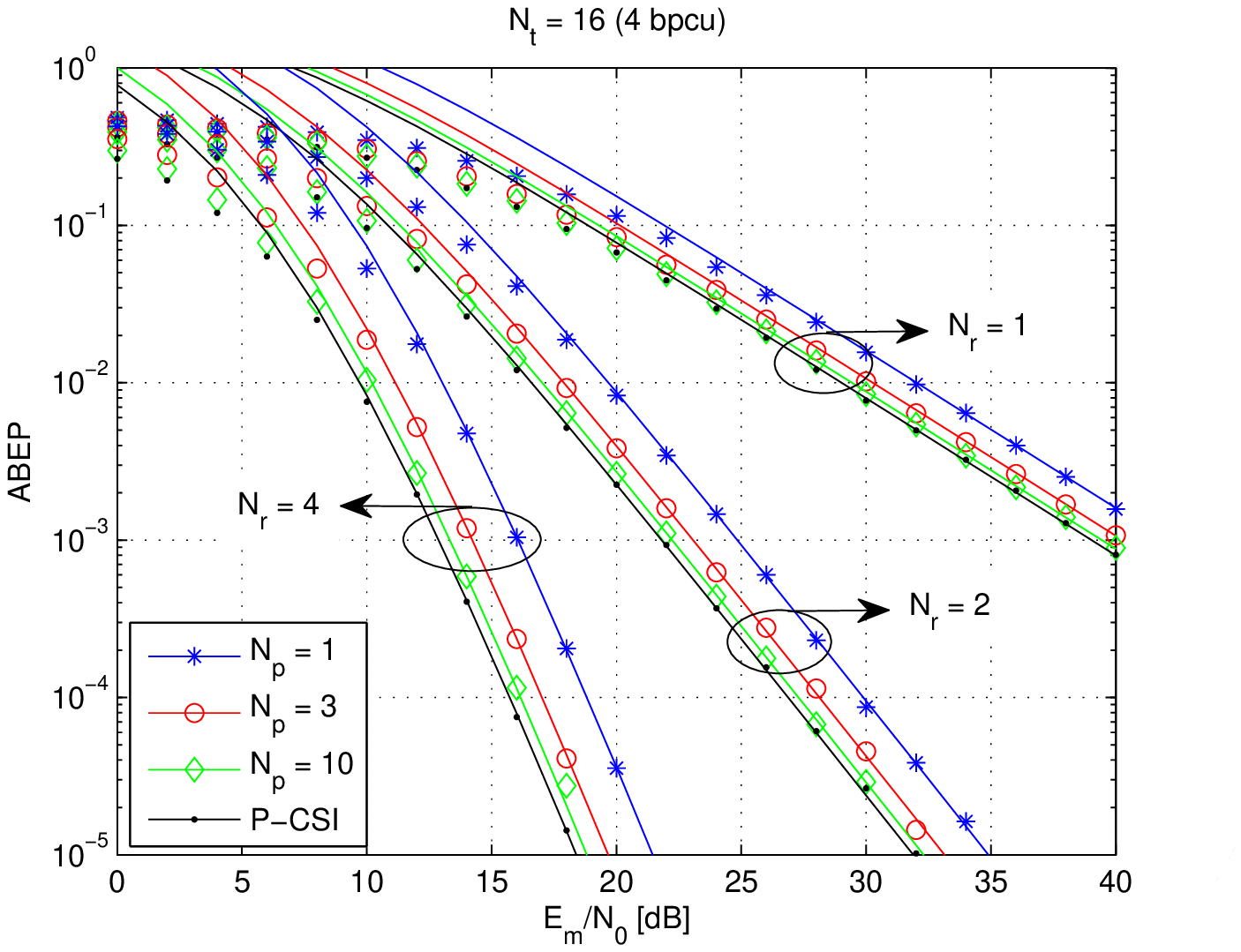}
\caption{ABEP of SSK modulation against $E_m/N_0$ for: i) $N_t = 16$ (4 bpcu); ii) $N_r = \left\{ {1,2,4} \right\}$; iii) $N_p  = \left\{
{1,3,10} \right\}$; and iv) P--CSI denotes the ABEP with no channel estimation errors. Solid lines show the analytical model and markers show Monte
Carlo simulations.} \label{Fig_SSK4bps}
\end{figure}
\begin{figure}[!t]
\centering
\includegraphics [width=\columnwidth] {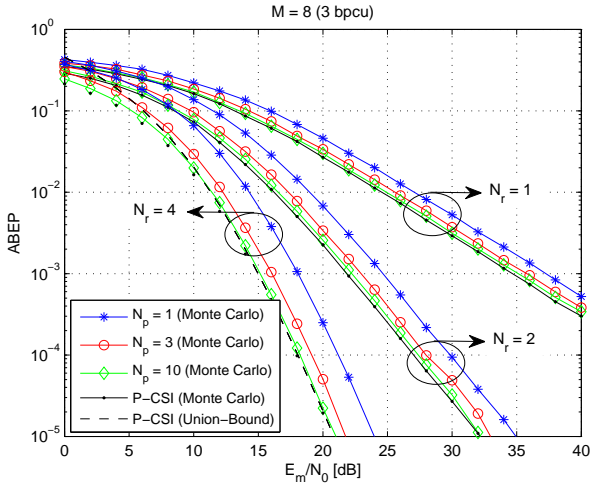}
\caption{ABEP of QAM against $E_m/N_0$ for: i) $M = 8$ (3 bpcu); ii) $N_r = \left\{ {1,2,4} \right\}$; iii) $N_p  = \left\{ {1,3,10} \right\}$;
and iv) P--CSI denotes the ABEP with no channel estimation errors. Solid lines with markers or just markers show Monte Carlo simulations.
Dashed lines show the union--bound computed from \cite{MDR_TVT} with no channel estimation errors at the receiver (P--CSI scenario). This
union--bound is shown only for a subset of curves in order to improve the readability of the figure, and avoid overlap among closely--spaced
curves.} \label{Fig_SingleAntennaQam3bps}
\end{figure}
\begin{figure}[!t]
\centering
\includegraphics [width=\columnwidth] {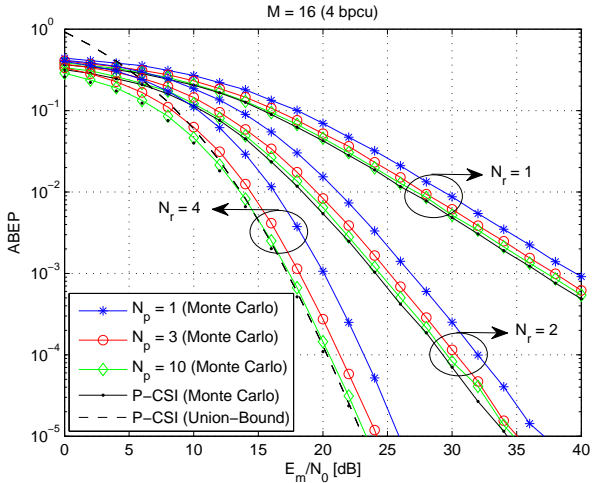}
\caption{ABEP of QAM against $E_m/N_0$ for: i) $M = 16$ (4 bpcu); ii) $N_r = \left\{ {1,2,4} \right\}$; iii) $N_p  = \left\{ {1,3,10}
\right\}$; and iv) P--CSI denotes the ABEP with no channel estimation errors. Solid lines with markers or just markers show Monte Carlo
simulations. Dashed lines show the union--bound computed from \cite{MDR_TVT} with no channel estimation errors at the receiver (P--CSI
scenario). This union--bound is shown only for a subset of curves in order to improve the readability of the figure, and avoid overlap among
closely--spaced curves.} \label{Fig_SingleAntennaQam4bps}
\end{figure}
\begin{figure}[!t]
\centering
\includegraphics [width=\columnwidth] {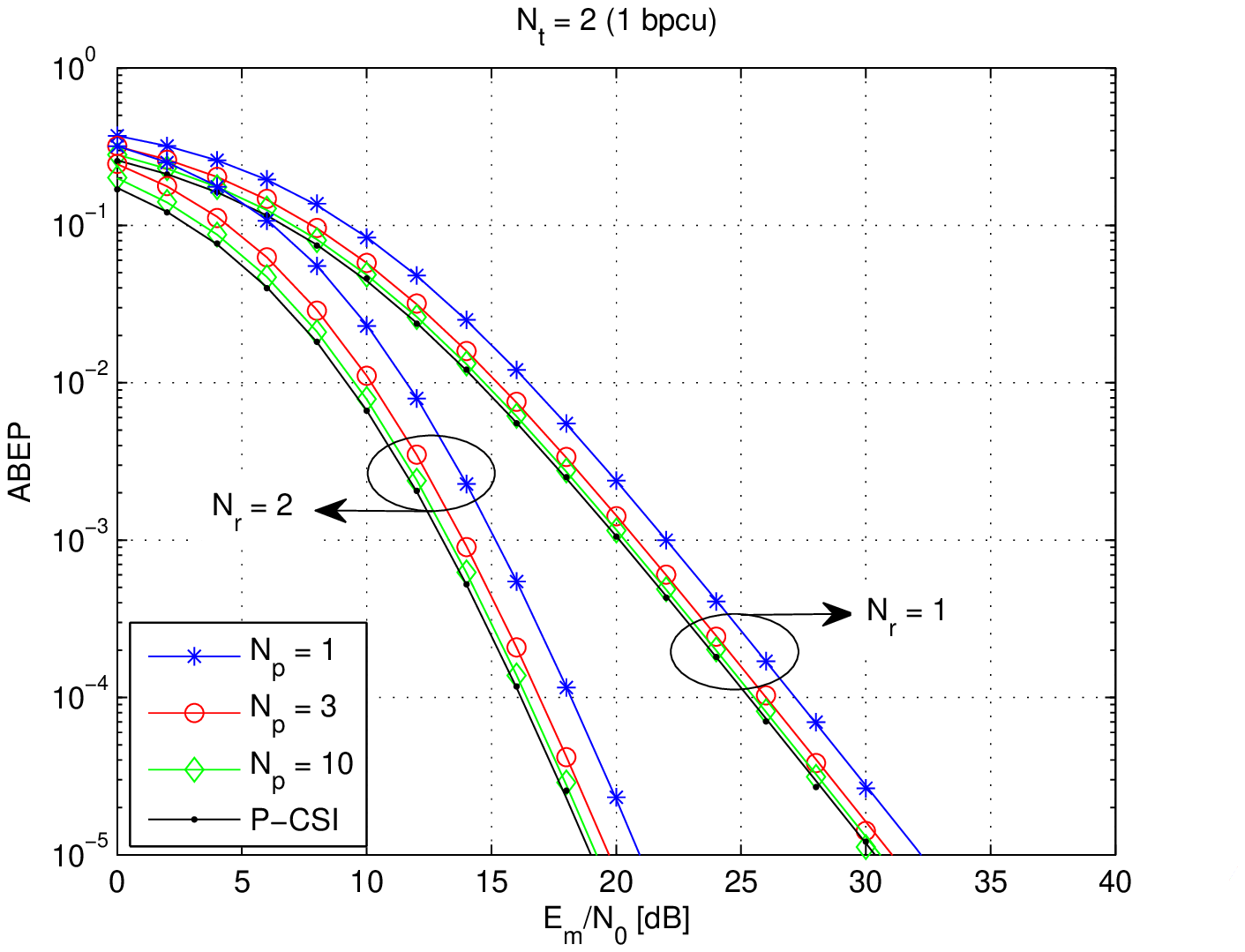}
\caption{ABEP of TOSD--SSK modulation against $E_m/N_0$ for: i) $N_t = 2$ (1 bpcu); ii) $N_r = \left\{ {1,2} \right\}$; iii) $N_p  = \left\{
{1,3,10} \right\}$; and iv) P--CSI denotes the ABEP with no channel estimation errors. Solid lines show the analytical model and markers show Monte
Carlo simulations.} \label{Fig_TOSD_SSK1bps}
\end{figure}
\begin{figure}[!t]
\centering
\includegraphics [width=\columnwidth] {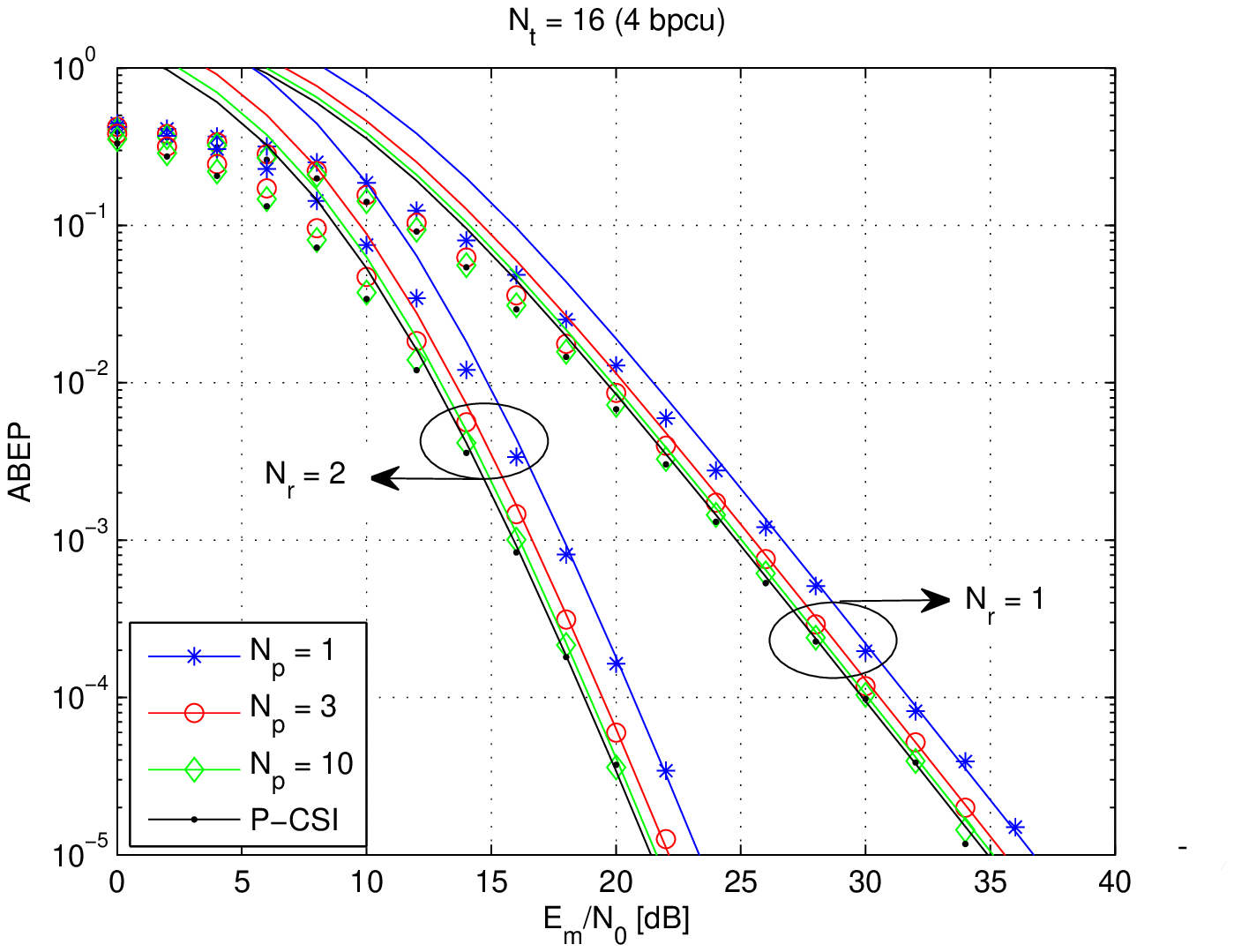}
\caption{ABEP of TOSD--SSK modulation against $E_m/N_0$ for: i) $N_t = 16$ (4 bpcu); ii) $N_r = \left\{ {1,2} \right\}$; iii) $N_p  = \left\{
{1,3,10} \right\}$; and iv) P--CSI denotes the ABEP with no channel estimation errors. Solid lines show the analytical model and markers show Monte
Carlo simulations.} \label{Fig_TOSD_SSK4bps}
\end{figure}
\begin{figure}[!t]
\centering
\includegraphics [width=\columnwidth] {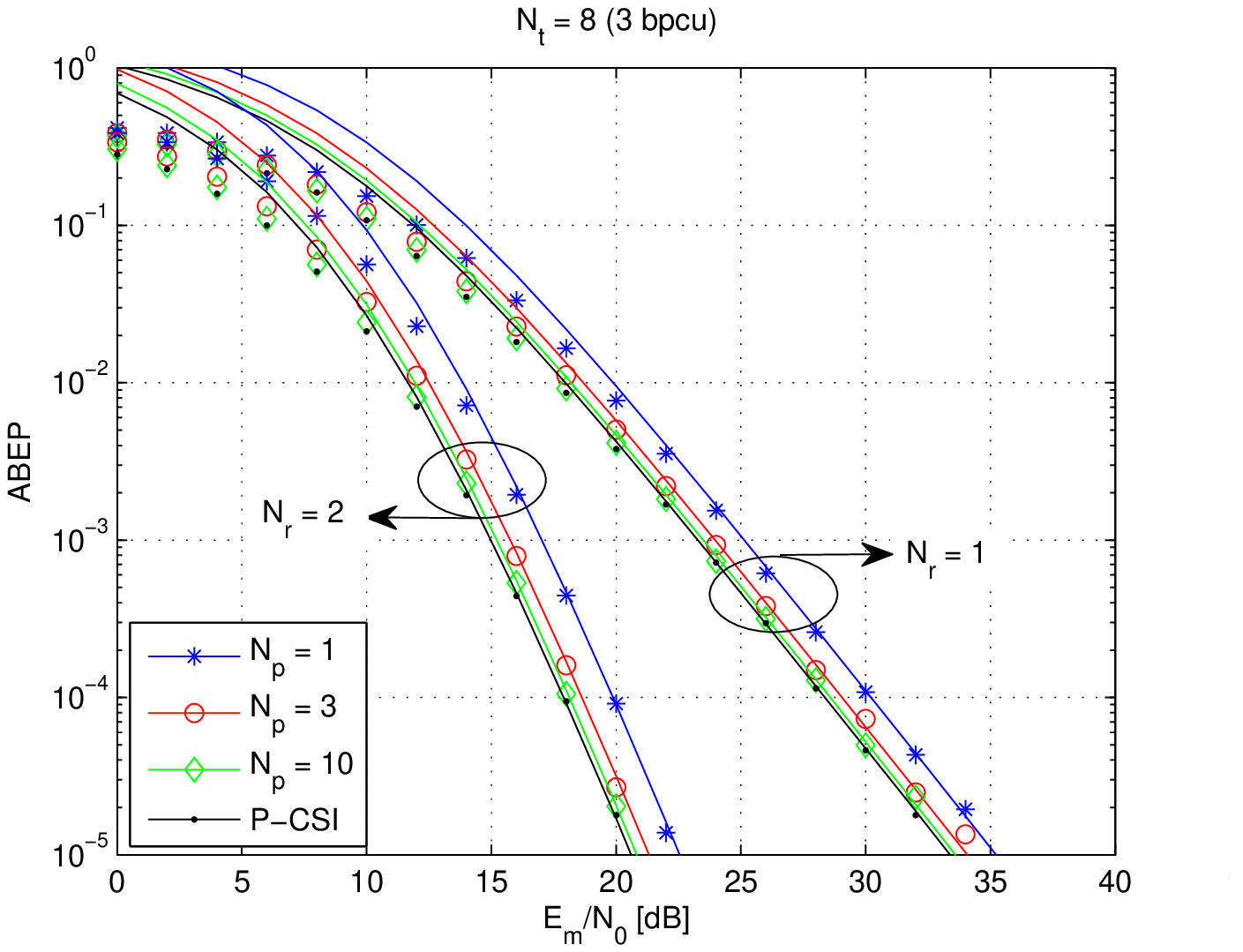}
\caption{ABEP of TOSD--SSK modulation against $E_m/N_0$ for: i) $N_t = 8$ (3 bpcu); ii) $N_r = \left\{ {1,2} \right\}$; iii) $N_p  = \left\{
{1,3,10} \right\}$; and iv) P--CSI denotes the ABEP with no channel estimation errors. Solid lines show the analytical model and markers show Monte
Carlo simulations.} \label{Fig_TOSD_SSK3bps} \vspace{-0.4cm}
\end{figure}
\begin{figure}[!t]
\centering
\includegraphics [width=\columnwidth] {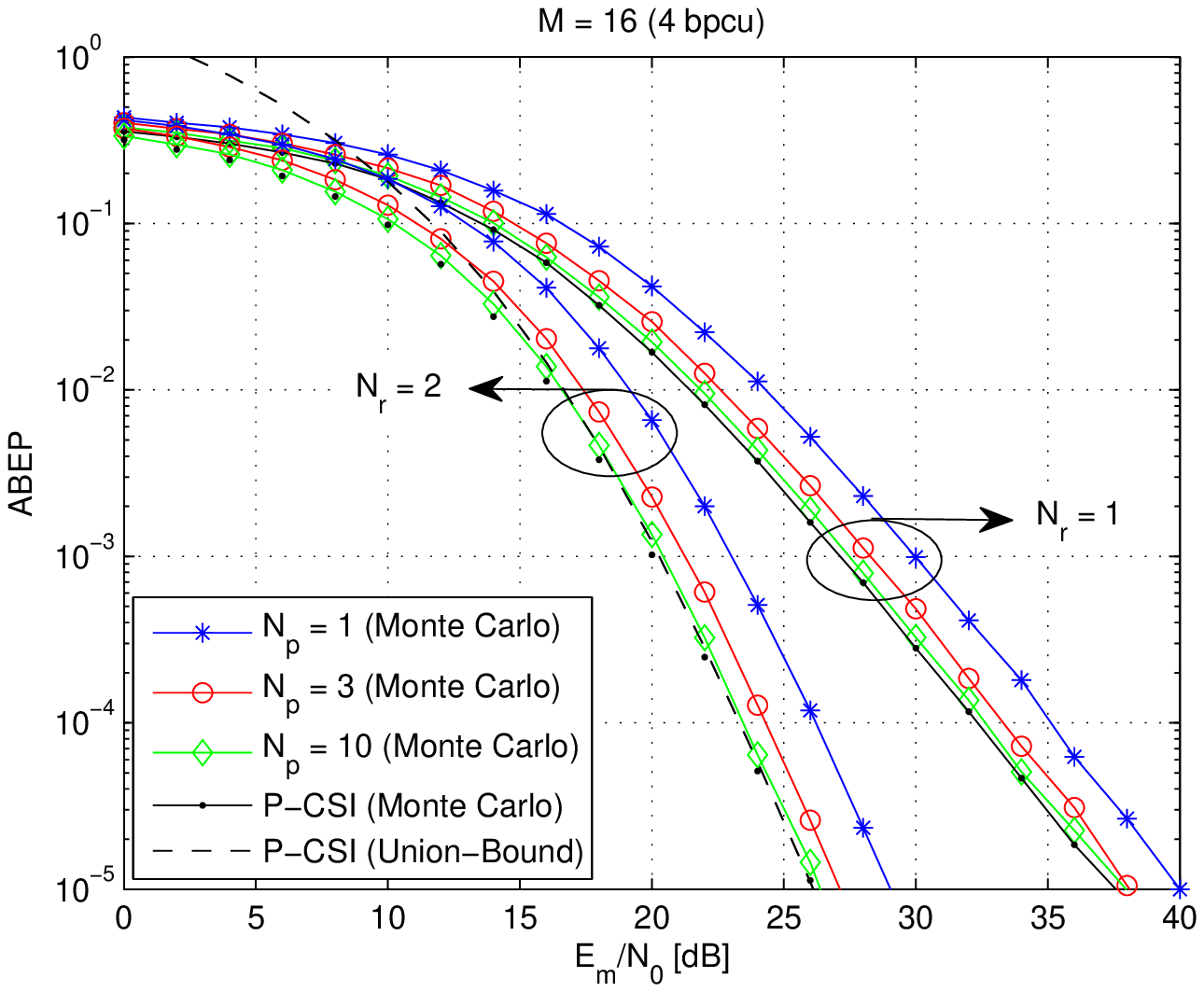}
\caption{ABEP of Alamouti scheme with QAM against $E_m/N_0$ for: i) $M = 16$ (4 bpcu); ii) $N_r = \left\{ {1,2} \right\}$; iii) $N_p = \left\{
{1,3,10} \right\}$; and iv) P--CSI denotes the ABEP with no channel estimation errors.  Solid lines with markers or just markers show Monte
Carlo simulations. Dashed lines show the union--bound computed from \cite{MDR_TVT} with no channel estimation errors at the receiver (P--CSI
scenario). This union--bound is shown only for a subset of curves in order to improve the readability of the figure, and avoid overlap among
closely--spaced curves.} \label{Fig_AlamoutiQam4bps}
\end{figure}

The results are shown in Figs. \ref{Fig_SSK1bps}--\ref{Fig_SingleAntennaQam4bps} for transmission technologies with no transmit--diversity gain
(SSK and QAM), and in Figs. \ref{Fig_TOSD_SSK1bps}--\ref{Fig_AlamoutiQam4bps} for transmission technologies with transmit--diversity gain
(TOSD--SSK and Alamouti). As far as SSK and TOSD--SSK modulations are concerned, we observe that: i) our analytical frameworks are very
accurate and asymptotically--tight for all the analyzed scenarios. In particular, as expected, they are exact for $N_t=2$; ii) there is no loss
of the diversity order in the presence of channel estimation errors. Only a loss of the coding gain can be observed for all MIMO setups; iii)
even though in space modulation the information is encoded into the CIRs, the performance degradation observed when reducing the number of
pilot pulses, $N_p$, is not very high, and the ABEP is very close to the P--CSI lower--bound, in the analyzed scenarios, for $N_p=10$; iv) SSK
and TOSD--SSK modulations have a SNR penalty, with respect to the P--CSI lower--bound, of approximately 3dB and 2dB when $N_p=1$, respectively;
v) the ABEP gets worse for increasing $N_t$, as a consequence of the increased size of the spatial--constellation diagram, and gets better for
increasing $N_r$, due to the receive--diversity gain; and vi) TOSD--SSK modulation significantly outperforms SSK modulation, due to the
transmit--diversity gain introduced by the orthogonal pulse shaping design.

As far as the performance comparison with single--antenna and Alamouti schemes is concerned, the following conclusions can be drawn (see Table
\ref{Tab_1} for numerical values): i) SSK modulation outperforms single--antenna QAM, in all the analyzed scenarios, for spectral efficiencies
greater than 2 bpcu and for $N_r > 1$. If $N_r = 1$, QAM always outperforms SSK modulation; ii) SSK and single--antenna QAM have almost the
same robustness to channel estimation errors, with a SNR penalty, with respect to the P--CSI lower--bound, of approximately 3dB when $N_p=1$;
iii) TOSD--SSK modulation outperforms the Alamouti scheme with QAM, in all the analyzed scenarios, for spectral efficiencies greater than 2
bpcu. In particular, unlike SSK modulation, TOSD--SSK modulation is superior to the Alamouti scheme with QAM for $N_r=1$ as well. This is due
to the transmit--diversity gain of TOSD--SSK modulation; iv) TOSD--SSK modulation is more robust to channel estimation errors than the Alamouti
scheme with QAM. A clear example can be observed in Table \ref{Tab_1} when $R = 2$ bpcu and $N_p=1$. In fact, the Alamouti scheme is superior
to TOSD--SSK modulation in the P--CSI scenario, but TOSD--SSK modulation provides better performance if $N_p=1$ and $N_r>1$. More in general,
Table \ref{Tab_1} shows that the Alamouti scheme with QAM has a SNR penalty, with respect to the P--CSI lower--bound, of approximately 3dB when
$N_p=1$, while TOSD--SSK modulation has a SNR penalty of only 2dB; and v) the performance gain of SSK and TOSD--SSK modulations with respect to
single--antenna and Alamouti schemes increases with $N_r$, because, as analytically proved in \cite{MDR_TVT}, space modulation takes much
better advantage of receive--diversity. The results shown in this section confirm that this trend is retained in the presence of
channel estimation errors as well.
In conclusion, SSK modulation is as robust as single--antenna systems to imperfect channel knowledge, and it provides better performance when
the target spectral efficiency is greater than 2 bpcu and $N_r>1$. On the other hand, TOSD--SSK modulation is more robust than the Alamouti
scheme to imperfect channel knowledge, and it provides better performance when the target spectral efficiency is greater than 2 bpcu. In all
the cases, the price to be paid for this performance improvement is the need of increasing the number of radiating elements $N_t$ at the
transmitter, while still retaining a single--RF chain and avoiding inter--antenna synchronization, which are beneficial for low--complexity
implementations \cite{Papadias_Aug2008}. This remark is somehow similar to \cite{Tarokh}, as far as the achievable transmit--diversity of STBCs
is concerned. Finally, it is worth emphasizing that the need of a large number of radiating elements seems not to be a critical bottleneck for
the development of the next generation cellular systems, as current research is moving towards the utilization of the millimeter--wave
frequency spectrum \cite{WCNC_2011__mmWaveTutorial}. In fact, in this band compact horn antenna--arrays with 48 elements and compact patch
antenna--arrays with more than 4 elements at the base station and at the mobile terminal, respectively, are currently being developed to
support multi--gigabit transmission rates \cite{GLOBE_2011__LargeMIMO}. Furthermore, SSK and TOSD--SSK seem to be well--suited low--complexity modulation schemes for the recently proposed ``massive MIMO'' paradigm \cite{Marzetta}, according to which unprecedent spectral efficiencies can be achieved in cellular networks by using antenna--arrays with very large (with tens or hundreds) active radiating elements.
\section{Conclusion} \label{Conclusion}
In this paper, we have analyzed the performance of space modulation when CSI is not perfectly known at the receiver. A very accurate and
general analytical framework has been proposed, and it has been shown that, unlike common belief, SSK modulation has the same robustness to
channel estimation errors as conventional modulation schemes, while TOSD--SSK modulation is less sensitive to channel estimation errors than
conventional modulations. Also, it has been shown that few pilot pulses are needed to achieve almost the same performance as the
P--CSI lower--bound, and that the performance gain, over state--of--the--art MIMO technologies, promised by space modulation is retained even
with imperfect channel knowledge. These results confirm the usefulness of space modulation in practical operating conditions, and, in
particular, the notable performance advantage of TOSD--SSK modulation, which provides transmit--diversity and is more robust
to channel estimation errors than conventional schemes, such as the Alamouti code.
\section*{Acknowledgment}
We gratefully acknowledge support from the European Union (PITN--GA--2010--264759, GREENET project) for this work. M. Di Renzo acknowledges
support of the Laboratory of Signals and Systems under the research project ``Jeunes Chercheurs''. D. De Leonardis and F. Graziosi
acknowledge the Italian Inter--University Consortium for Telecommunications (CNIT) under the research grant ``Space Modulation for MIMO
Systems'', and Micron Technology under the Ph.D. fellowships program awarded to the University of L'Aquila. H. Haas acknowledges the EPSRC under grant EP/G011788/1 for partially funding this work.
\appendices
\begin{table*}[!t]
\renewcommand{\arraystretch}{1.2}
\caption{Required $E_m/N_0$ (${\rm{dB}}$) to get ${\rm{ABEP}}= 10^{-4}$ for all scenarios except SSK modulation and single--antenna QAM if
$N_r=1$, for which the $E_m/N_0$ (${\rm{dB}}$) to get ${\rm{ABEP}}= 10^{-2}$ is shown. For SSK modulation and single--antenna QAM, each row
shows the $E_m/N_0$ (${\rm{dB}}$) for $N_r=1$ / $N_r=2$ / $N_r=4$. For SSK modulation and the Alamouti scheme with QAM, each row shows the
$E_m/N_0$ (${\rm{dB}}$) for $N_r=1$ / $N_r=2$. The values have an error approximately equal to $\pm 0.1{\rm{dB}}$.} \label{Tab_1}
\begin{center}
\begin{tabular}{|c||c|c|c|c|}
\hline
\multicolumn{5}{|c|} {SSK} \\
\hline
Rate & $N_p=1$ & $N_p=3$ & $N_p=10$ & ${\rm{P-CSI}}$ \\
\hline
1 bpcu & 22.9 / 25.3 / 16.2 & 21.1 / 23.5 / 14.5 & 20.3 / 22.7 / 13.6 & 19.9 / 22.3 / 13.2 \\
\hline
2 bpcu & 26 / 26.8 / 17 & 24.2 / 25.1 / 15.4 & 23.4 / 24.3 / 14.5 & 23 / 23.8 / 14 \\
\hline
3 bpcu & 29 / 28.4 / 17.9 & 27.3 / 26.6 / 16.2 & 26.4 / 25.8 / 15.3 & 26 / 25.4 / 14.9 \\
\hline
4 bpcu & 32 / 29.9 / 18.7 & 30.3 / 28.1 / 17 & 29.5 / 27.3 / 16.2 & 29 / 26.9 / 15.7 \\
\hline \hline
\multicolumn{5}{|c|} {Single--Antenna QAM} \\
\hline
Rate & $N_p=1$ & $N_p=3$ & $N_p=10$ & ${\rm{P-CSI}}$ \\
\hline
1 bpcu & 19.8 / 22.3 / 13.1  &  18.2 / 20.6 / 11.5 & 17.2 / 19.7 / 10.6 & 16.8 / 19.3 / 10.1 \\
\hline
2 bpcu & 22.7 / 25.2 / 16.2 & 21.1 / 23.5 / 14.5 & 20.3 / 22.7 / 13.6 & 19.9 / 22.2 / 13.2 \\
\hline
3 bpcu & 27.5 / 29.9 / 21.1 & 25.6 / 28 / 19.1  &  24.9 / 27.4 / 18.2 & 24.6 / 27 / 18.1 \\
\hline
4 bpcu & 29.6 / 32 / 23.3  &  27.8 / 30.1 / 21.3 & 27.1 / 29.6 / 20.5 & 26.8 / 29.1 / 20.3 \\
\hline \hline
\multicolumn{5}{|c|} {TOSD--SSK} \\
\hline
Rate & $N_p=1$ & $N_p=3$ & $N_p=10$ & ${\rm{P-CSI}}$ \\
\hline
1 bpcu & 27.2 / 18.2 & 26 / 16.9  & 25.5 / 16.4 & 25.3 / 16.2 \\
\hline
2 bpcu & 28.7 / 19 &  27.5 / 17.8 & 27 / 17.3  & 26.8 / 17 \\
\hline
3 bpcu & 30.2 / 19.8 & 29 / 18.6 &  28.5 / 18.2 & 28.4 / 17.8 \\
\hline
4 bpcu & 31.9 / 20.7 & 30.5 / 19.4 & 30.1 / 18.9 & 29.9 / 18.7 \\
\hline \hline
\multicolumn{5}{|c|} {Alamouti QAM} \\
\hline
Rate & $N_p=1$ & $N_p=3$ & $N_p=10$ & ${\rm{P-CSI}}$ \\
\hline
1 bpcu & 25.3 / 16.2 & 23.5 / 14.5 & 22.8 / 13.5 & 22.3 / 13.2 \\
\hline
2 bpcu & 28.4 / 19.3 & 26.5 / 17.5 & 25.7 / 16.6 & 25.4 / 16.3 \\
\hline
3 bpcu & 32.9 / 24  & 31.4 / 22.2 & 30.4 / 21.3 & 30 / 21 \\
\hline
4 bpcu & 35.2 / 26.2 & 33.3 / 24.3 & 32.6 / 23.5 & 32.3 / 23.3 \\
\hline
\end{tabular}
\end{center}
\end{table*}
\begin{figure*}[!t]
\setcounter{equation}{34}
\begin{equation}
\label{Eq_App1} \left\{ \begin{array}{l}
 w_{t_1 } \left( \xi  \right) = \left( { - \frac{4}{{\sqrt {165} }}} \right)p_1 \left( \xi  \right) + \left( {\frac{{\sqrt {4 - 2\sqrt 2 } }}{4}} \right)p_2 \left( \xi  \right) + \left( {\sqrt {\frac{{11}}{{30}}} } \right)p_3 \left( \xi  \right) + \left( { - \frac{{\sqrt {4 + 2\sqrt 2 } }}{4}} \right)p_4 \left( \xi  \right) + \left( { - \frac{2}{{\sqrt {110} }}} \right)p_5 \left( \xi  \right) \\
 w_{t_2 } \left( \xi  \right) = \left( { - \frac{4}{{\sqrt {165} }}} \right)p_1 \left( \xi  \right) + \left( { - \frac{{\sqrt {4 - 2\sqrt 2 } }}{4}} \right)p_2 \left( \xi  \right) + \left( {\sqrt {\frac{{11}}{{30}}} } \right)p_3 \left( \xi  \right) + \left( {\frac{{\sqrt {4 + 2\sqrt 2 } }}{4}} \right)p_4 \left( \xi  \right) + \left( { - \frac{2}{{\sqrt {110} }}} \right)p_5 \left( \xi  \right) \\
 w_{t_3 } \left( \xi  \right) = \left( {\sqrt {\frac{3}{{22}}} } \right)p_1 \left( \xi  \right) + \left( { - \frac{{\sqrt {4 + 2\sqrt 2 } }}{4}} \right)p_2 \left( \xi  \right) + \left( 0 \right)p_3 \left( \xi  \right) + \left( {\frac{{\sqrt {4 - 2\sqrt 2 } }}{4}} \right)p_4 \left( \xi  \right) + \left( { - \frac{2}{{\sqrt {11} }}} \right)p_5 \left( \xi  \right) \\
 w_{t_4 } \left( \xi  \right) = \left( {\sqrt {\frac{3}{{22}}} } \right)p_1 \left( \xi  \right) + \left( {\frac{{\sqrt {4 + 2\sqrt 2 } }}{4}} \right)p_2 \left( \xi  \right) + \left( 0 \right)p_3 \left( \xi  \right) + \left( { - \frac{{\sqrt {4 - 2\sqrt 2 } }}{4}} \right)p_4 \left( \xi  \right) + \left( { - \frac{2}{{\sqrt {11} }}} \right)p_5 \left( \xi  \right) \\
 \end{array} \right.
\end{equation}
\normalsize \hrulefill \vspace*{-5pt}
\end{figure*}
\begin{figure*}[!t]
\setcounter{equation}{35}
\begin{equation}
\label{Eq_App3} \left\{ \begin{array}{l}
 W_{t_1 } \left( \omega  \right) = \left( { - \frac{4}{{\sqrt {165} }}} \right)P_1 \left( \omega  \right) + \left( {\frac{{\sqrt {4 - 2\sqrt 2 } }}{4}} \right)P_2 \left( \omega  \right) + \left( {\sqrt {\frac{{11}}{{30}}} } \right)P_3 \left( v \right) + \left( { - \frac{{\sqrt {4 + 2\sqrt 2 } }}{4}} \right)P_4 \left( \omega  \right) + \left( { - \frac{2}{{\sqrt {110} }}} \right)P_5 \left( \omega  \right) \\
 W_{t_2 } \left( \omega  \right) = \left( { - \frac{4}{{\sqrt {165} }}} \right)P_1 \left( \omega  \right) + \left( { - \frac{{\sqrt {4 - 2\sqrt 2 } }}{4}} \right)P_2 \left( \omega  \right) + \left( {\sqrt {\frac{{11}}{{30}}} } \right)P_3 \left( \omega  \right) + \left( {\frac{{\sqrt {4 + 2\sqrt 2 } }}{4}} \right)P_4 \left( \omega  \right) + \left( { - \frac{2}{{\sqrt {110} }}} \right)P_5 \left( \omega  \right) \\
 W_{t_3 } \left( \omega  \right) = \left( {\sqrt {\frac{3}{{22}}} } \right)P_1 \left( \omega  \right) + \left( { - \frac{{\sqrt {4 + 2\sqrt 2 } }}{4}} \right)P_2 \left( \omega  \right) + \left( 0 \right)P_3 \left( \omega  \right) + \left( {\frac{{\sqrt {4 - 2\sqrt 2 } }}{4}} \right)P_4 \left( \omega  \right) + \left( { - \frac{2}{{\sqrt {11} }}} \right)P_5 \left( \omega  \right) \\
 W_{t_4 } \left( \omega  \right) = \left( {\sqrt {\frac{3}{{22}}} } \right)P_1 \left( \omega  \right) + \left( {\frac{{\sqrt {4 + 2\sqrt 2 } }}{4}} \right)P_2 \left( \omega  \right) + \left( 0 \right)P_3 \left( \omega  \right) + \left( { - \frac{{\sqrt {4 - 2\sqrt 2 } }}{4}} \right)P_4 \left( \omega  \right) + \left( { - \frac{2}{{\sqrt {11} }}} \right)P_5 \left( \omega  \right) \\
 \end{array} \right.
\end{equation}
\normalsize \hrulefill \vspace*{-10pt}
\end{figure*}
\section{Orthogonal Shaping Filters for $N_t=4$} \label{App}
In this appendix, we show an example of orthogonal shaping filters that can be used for TOSD--SSK modulation. Without loss of generality we consider the case study with $N_t=4$, but the procedure can be generalized to larger antenna--arrays.

More specifically, we consider the procedure described in \cite{TCOM_PSM}, which allows us to generate orthogonal shaping filters with the same time--duration and bandwidth. Similar techniques are available in \cite{Parr_BW}, \cite{Giannakis_BW}. From \cite{TCOM_PSM}, we can obtain the four orthogonal impulse (time) responses shown in (\ref{Eq_App1}) at the bottom of the previous page, as well as the four related frequency responses (Fourier transform) $P\left( \omega  \right) = \left( {{1 \mathord{\left/ {\vphantom {1 {\sqrt {2\pi } }}} \right. \kern-\nulldelimiterspace} {\sqrt {2\pi } }}} \right)\int\nolimits_{ - \infty }^{ + \infty } {p\left( \xi  \right)\exp \left( { - j\omega \xi } \right)d\xi }$ shown in (\ref{Eq_App3}) at the bottom of the previous page too, where we have defined:
\setcounter{equation}{36}
\begin{equation}
\label{Eq_App2} \left\{ \begin{array}{l}
 p_1 \left( \xi  \right) = \left( {\frac{1}{{\sqrt {\sqrt \pi  } }}} \right)\exp \left[ { - \frac{1}{2}\left( {\frac{\xi }{{t_0 }}} \right)^2 } \right] \\
 p_2 \left( \xi  \right) = \left( {\frac{{2\xi }}{{\sqrt {2\sqrt \pi  } }}} \right)\exp \left[ { - \frac{1}{2}\left( {\frac{\xi }{{t_0 }}} \right)^2 } \right] \\
 p_3 \left( \xi  \right) = \left( {\frac{{4\xi ^2  - 2}}{{\sqrt {8\sqrt \pi  } }}} \right)\exp \left[ { - \frac{1}{2}\left( {\frac{\xi }{{t_0 }}} \right)^2 } \right] \\
 p_4 \left( \xi  \right) = \left( {\frac{{8\xi ^3  - 12\xi }}{{\sqrt {48\sqrt \pi  } }}} \right)\exp \left[ { - \frac{1}{2}\left( {\frac{\xi }{{t_0 }}} \right)^2 } \right] \\
 p_5 \left( \xi  \right) = \left( {\frac{{16\xi ^4  - 48\xi ^2  + 12}}{{\sqrt {384\sqrt \pi  } }}} \right)\exp \left[ { - \frac{1}{2}\left( {\frac{\xi }{{t_0 }}} \right)^2 } \right] \\
 \end{array} \right.
\end{equation}
\noindent and:
\setcounter{equation}{37}
\begin{equation}
\label{Eq_App4}
 \left\{ \begin{array}{l}
 P_1 \left( \omega  \right) = \left( {\frac{{t_0 }}{{\sqrt {\sqrt \pi  } }}} \right)\exp \left[ { - \frac{1}{2}\left( {t_0 \omega } \right)^2 } \right] \\
 P_2 \left( \omega  \right) = \left( {\frac{{2jt_0^2 \omega }}{{\sqrt {2\sqrt \pi  } }}} \right)\exp \left[ { - \frac{1}{2}\left( {t_0 \omega } \right)^2 } \right] \\
 P_3 \left( \omega  \right) = \left( {\frac{{2t_0  - 4t_0^3 \omega ^2 }}{{\sqrt {8\sqrt \pi  } }}} \right)\exp \left[ { - \frac{1}{2}\left( {t_0 \omega } \right)^2 } \right] \\
 P_4 \left( \omega  \right) = \left( {\frac{{12jt_0^2 \omega  - 8jt_0^4 \omega ^3 }}{{\sqrt {48\sqrt \pi  } }}} \right)\exp \left[ { - \frac{1}{2}\left( {t_0 \omega } \right)^2 } \right] \\
 P_5 \left( \omega  \right) = \left( {\frac{{12t_0  - 48t_0^3 \omega ^2  + 16t_0^5 \omega ^4 }}{{\sqrt {384\sqrt \pi  } }}} \right)\exp \left[ { - \frac{1}{2}\left( {t_0 \omega } \right)^2 } \right] \\
 \end{array} \right.
\end{equation}

Finally, we mention that, by adjusting the form factor $t_0$, the bandwidth can be arbitrarily chosen, and both narrow-- and
wide--band communication systems can be considered.
\begin{biography}[{\includegraphics[width=1in,height=1.25in,clip,keepaspectratio]{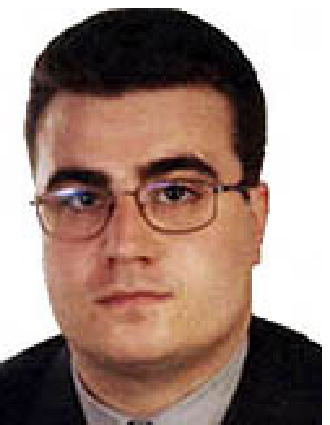}}]{Marco Di Renzo}
(SM'05--AM'07--M'09) was born in L'Aquila, Italy, in 1978. He received the Laurea (cum laude) and the Ph.D. degrees in Electrical and
Information Engineering from the Department of Electrical and Information Engineering, University of L'Aquila, Italy, in April 2003 and in January
2007, respectively.

From August 2002 to January 2008, he was with the Center of Excellence for Research DEWS, University of L'Aquila, Italy.
From February 2008 to April 2009, he was a Research Associate with the Telecommunications Technological Center of Catalonia (CTTC), Barcelona,
Spain. From May 2009 to December 2009, he was an EPSRC Research Fellow with the Institute for Digital Communications (IDCOM), The University of
Edinburgh, Edinburgh, United Kingdom (UK).

Since January 2010, he has been a Tenured Researcher (``Charg\'e de Recherche Titulaire'') with the French National Center for Scientific
Research (CNRS), as well as a research staff member of the Laboratory of Signals and Systems (L2S), a joint research laboratory of the CNRS,
the \'Ecole Sup\'erieure d'\'Electricit\'e (SUP\'ELEC), and the University of Paris--Sud XI, Paris, France. His main research interests are in
the area of wireless communications theory, signal processing, and information theory.

Dr. Di Renzo is the recipient of the special mention for the outstanding five--year (1997--2003) academic career, University of L'Aquila, Italy;
the THALES Communications fellowship for doctoral studies (2003--2006), University of L'Aquila, Italy; and the Torres Quevedo award for his research on ultra wide band systems and cooperative localization for wireless networks (2008--2009), Ministry of Science and Innovation, Spain.
\end{biography}
\begin{biography}[{\includegraphics[width=1in,height=1.25in,clip,keepaspectratio]{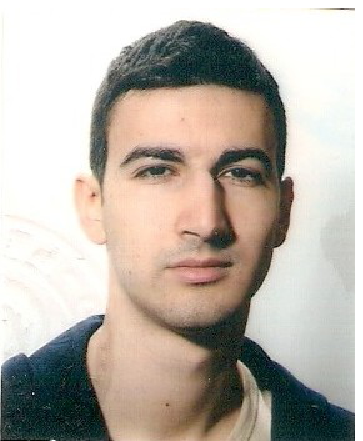}}]{Dario De Leonardis}
was born in Atri, Italy, in 1984. He received the Bachelor and the Master degrees in Telecommunications Engineering from the Department of
Electrical and Information Engineering, University of L'Aquila, Italy, in October 2006 and in May 2009, respectively.

From September 2008 to March 2009, he was a Visiting ERASMUS student at the Telecommunications Technological Center of Catalonia (CTTC),
Barcelona, Spain, where he conducted research for his Master thesis project on low--complexity receiver design for ultra wide band wireless
systems. In April 2010, he received a personal research grant from the National Inter--University Consortium for Telecommunications
(CNIT), Italy, to conduct research on space modulation for multiple--input--multiple--output wireless systems, and he was affiliated with the
Department of Electrical and Information Engineering and the Center of Excellence for Research DEWS, University of L'Aquila, Italy. Since
December 2010, he has been a Ph.D. candidate in the same institution. His main research interests are in the area of wireless communications.
\end{biography}
\begin{biography}{Fabio Graziosi}
(S'96–-M'97) was born in L'Aquila, Italy, in 1968. He received the Laurea degree (cum laude) and the Ph.D. degree in electronic engineering
from the University of L'Aquila, Italy, in 1993 and in 1997, respectively.

Since February 1997, he has been with the Department of Electrical Engineering, University of L'Aquila, where he is currently an Associate
Professor. He is a member of the Executive Committee, Center of Excellence Design methodologies for Embedded controllers, Wireless interconnect
and System-on-chip (DEWS), University of L'Aquila, and the Executive Committee, Consorzio Nazionale Interuniversitario per le Telecomunicazioni
(CNIT). He is also the Chairman of the Board of Directors of WEST Aquila s.r.l., a spin--off R\&D company of the University of L'Aquila, and the
Center of Excellence DEWS. He is involved in major national and European research programs in the field of wireless systems and he has been a
reviewer for major technical journals and international conferences in communications. He also serves as Technical Program Committee (TPC)
member and Session Chairman of several international conferences in communications. His current research interests are mainly focused on
wireless communication systems with emphasis on wireless sensor networks, ultra wide band communication techniques, cognitive radio, and
cooperative communications.
\end{biography}
\begin{biography}[{\includegraphics[width=1in,height=1.25in,clip,keepaspectratio]{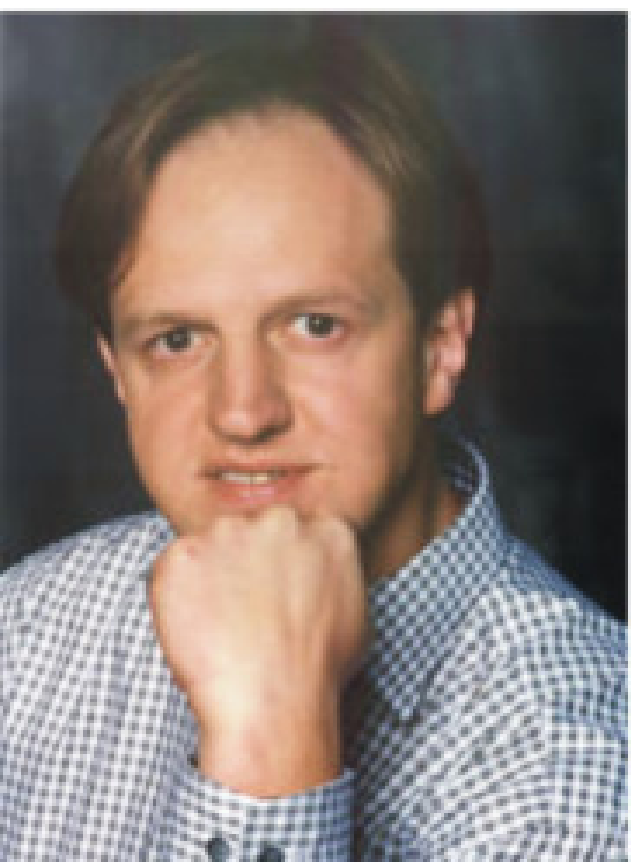}}]{Harald Haas}
(SM'98--AM'00--M'03) holds the Chair of Mobile Communications in the Institute for Digital Communications (IDCOM) at the University of Edinburgh. His
main research interests are in the areas of wireless system design and analysis as well as digital signal processing, with a particular focus on interference coordination in wireless networks, spatial modulation and optical wireless communication.

Professor Haas holds more than 15 patents. He has published more than 50 journal papers including a Science Article and more than 140 peer--reviewed conference papers. Nine of his papers are invited papers. He has co--authored a book entitled ``Next Generation Mobile Access Technologies: Implementing TDD'' with Cambridge University Press. Since 2007, he has been a Regular High Level Visiting Scientist supported by the Chinese ``111 program'' at Beijing University of Posts and Telecommunications (BUPT). He was an invited speaker at the TED Global conference 2011. He has been shortlisted for the World Technology Award for communications technology (individual) 2011.
\end{biography}
\end{document}